\documentclass{ws-procs961x669}            
\begin{document}
\title{Gravitational Lensing by Charged Accelerating Black Holes}

\author{Torben C. Frost}

\address{ZARM, University of Bremen, 28359 Bremen, Germany and Institute for Theoretical Physics, Leibniz University Hannover, 30167 Hannover, Germany\\
E-mail: torben.frost@zarm.uni-bremen.de}

\begin{abstract}
Current astrophysical observations show that on large scale the Universe is electrically
neutral. However, locally this may be quite different. Black holes enveloped by
a plasma in the presence of a strong magnetic field may have acquired a significant
electric charge. We can also expect that some of these charged black holes are moving.
Consequently to describe them we need spacetime metrics describing moving black
holes. In general relativity such a solution is given by the charged C-de Sitter-metric.
In this article we will assume that it can be used to describe moving charged black
holes. We will investigate how to observe the electric charge using gravitational
lensing. First we will use elliptic integrals and functions to solve the geodesic
equations. Then we will derive lens equation, travel time and redshift. We will
discuss the impact of the electric charge on these observables and potential limitations
for its observation.
\end{abstract}

\keywords{Black Holes; Gravitational Lensing}

\bodymatter

\section{Introduction}
X-ray and gravitational wave observations indicate that stellar mass black holes
are widely distributed in our galaxy.\cite{CorralSantana2016,Abbott2021} In addition,
the motion of gas and stars\cite{Gueltekin2012,GravCollab2020} around galactic
centres as well as the recent observation of the shadow of the compact object in
the centre of the galaxy M87\cite{EHTCollaboration2019} indicate that in the
centre of most galaxies we can find at least one supermassive black hole (SMBH).
While the resolution of these observations are not yet precise enough to exclude
all alternative theories of gravity or different types of compact objects all
observational features indicate that these compact objects can be described either
by the Schwarzschild metric or the Kerr metric.\\
Concurrent astrophysical observations show that our Universe is electrically neutral.
This indicates that all astrophysical objects, among them black holes, do not carry
a significant electric charge. However, if a black hole is surrounded by a plasma
with sufficiently large magnetic fields, the black hole may acquire a significant
electric charge.\cite{Castellanos2018} In general relativity charged black holes
that are at rest and do not carry a spin are described by the Reissner-Nordstr\"{o}m
metric, or, if we want to take the cosmological expansion into account, by the
Reissner-Nordstr\"{o}m-de Sitter metric. In the case that we consider real astrophysical
environments it is very unlikely that black holes are at rest. Instead it is more
realistic to assume that they constantly undergo accelerated motion. Consequently
we need spacetimes describing accelerating black holes to describe them accurately.
In the framework of general relativity a family of such solutions is described by
the charged C-de Sitter metric.\cite{Griffiths2009} The metric is an exact
solution to Einstein's electrovacuum field equations with cosmological constant.
It generalises the Reissner-Nordstr\"{o}m-de Sitter metric by introducing the acceleration
parameter $\alpha$ in addition to the mass parameter $m$, the electric charge $e$
and the cosmological constant $\Lambda$. The charged C-de Sitter metric is
axisymmetric and static and describes a charged accelerating black hole with
cosmological constant. The acceleration of the black hole results from conical
singularities on the axes, which are commonly interpreted as a string pulling the
black hole and a strut pushing the black hole. These conical singularities make
the spacetime on a first view appear unphysical. However, it may still serve
as an approximation for black holes undergoing accelerated motion.
Gravitational lensing in the C-metric only attracted attention relatively recently.
Grenzebach, Perlick and L\"ammerzahl\cite{Grenzebach2015a,Grenzebach2016} investigated
the shadow of accelerating black holes in the whole Pleba\'{n}ski-Demia\'{n}ski
family with acceleration.\cite{Plebanski1976} Sharif and Iftikhar\cite{Sharif2016}
calculated the deflection angle for light rays in the equatorial plane for accelerating
Kerr-NUT black holes. The method they used explicitly assumes that the light rays
are located in the equatorial plane of the spacetime. However, Alrais Alawadi, Batic
and Nowakowski\cite{AlraisAlawadi2021} and subsequently Frost and Perlick\cite{Frost2021}
demonstrated that the C-metric possesses a photon cone $\vartheta_{\mathrm{ph}}\neq\pi/2$
and therefore the method of Sharif and Iftikhar cannot be directly applied. Alrais
Alawadi, Batic and Nowakowski\cite{AlraisAlawadi2021} calculated the deflection
angle on the photon cone. Only a short time later Frost and Perlick\cite{Frost2021}
investigated gravitational lensing in the C-metric. In their work Frost and Perlick\cite{Frost2021}
used the canonical form of the elliptic integrals and Jacobi's
elliptic functions to solve the geodesic equations. Then, following the apporach
of Grenzebach, Perlick and L\"{a}mmerzahl,\cite{Grenzebach2015a} they fixed an
observer in the region of outer communication and introduced an orthonormal tetrad
to relate latitude and longitude on the observer's celestial sphere to the constants
of motion of the light rays. Using the conventions in Bohn et al.\cite{Bohn2015}
they formulated a lens equation, they derived the redshift and the travel time of individual
light rays on the celestial sphere of the observer.
In this article we will extend their approach to the charged C-de Sitter metric.
For this purpose the remainder of this article is structured as follows. As not
all readers may be familiar with the charged C-de Sitter metric we will provide
a short overview of its physical properties in Section \ref{sec:Metric}. In
Section \ref{sec:EoM} we will demonstrate how to solve the equations of motion.
In Section \ref{sec:Lensing} we will first introduce the orthonormal tetrad and
show how to parameterise the light rays using the angles on the observer's
celestial sphere. Then we will introduce the lens map, formulate the lens equation,
and calculate the redshift and the travel time of light rays. In Section \ref{sec:Conc}
we will shortly summarise our results and conclusions. Throughout the whole article
we will use geometric units with $c=G=1$. The metric signature is chosen as
$\left(-,+,+,+\right)$.

\section{The Charged C-de Sitter-Metric}\label{sec:Metric}
\begin{figure}[ht]\label{fig:horizons}
  \centering
		\includegraphics[width=\textwidth]{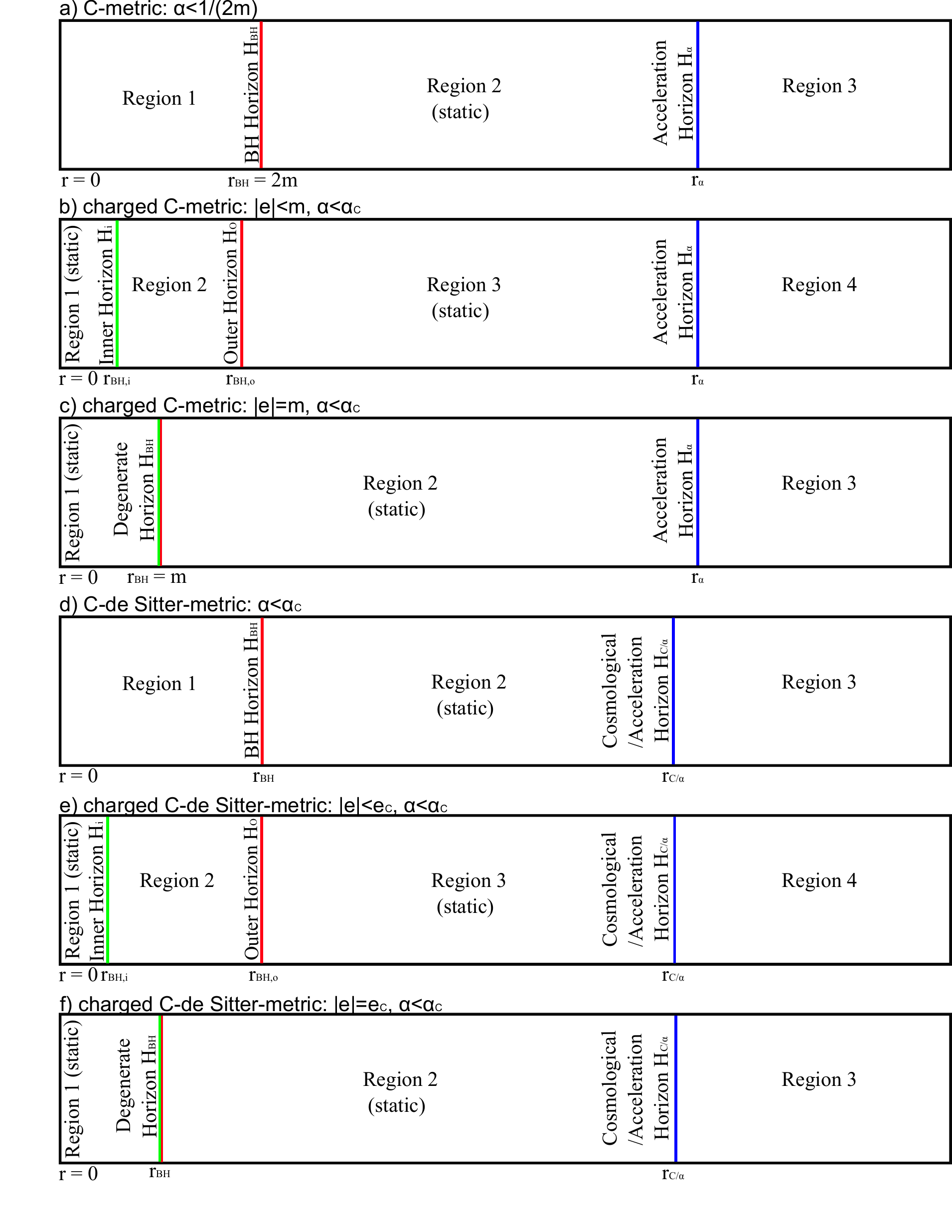}
	\caption{Position of the curvature singularity at $r=0$ and the coordinate
  singularities in a) the C-metric, the charged C-metric with b) $\left|e\right|<m$
  and c) $\left|e\right|=m$, d) the C-de Sitter metric and the charged C-de Sitter
  metric with e) $\left|e\right|<e_{\mathrm{C}}$ and f) $\left|e\right|=e_{\mathrm{C}}$.
  Note that the angular coordinates are suppressed and other singularities
  are not shown.}
\end{figure}
The charged C-de Sitter metric is a solution to Einstein's electrovacuum field equations
with cosmological constant. It belongs to the Pleba\'{n}ski-Demia\'{n}ski family
of spacetimes of Petrov type D\cite{Plebanski1976} and is axisymmetric and static.
The spacetime is characterised by four parameters: the mass parameter $m$, the
electric charge $e$, the cosmological constant $\Lambda$, and the acceleration
parameter $\alpha$. Its line element reads\cite{Griffiths2009}
\begin{equation}\label{eq:line}
g_{\mu\nu}\mathrm{d}x^{\mu}\mathrm{d}x^{\nu}=\frac{1}{\Omega(r,\vartheta)^2}\left(-Q(r)\mathrm{d}t^2+\frac{\mathrm{d}r^2}{Q(r)}+\frac{r^2\mathrm{d}\vartheta^2}{P(\vartheta)}+r^2\sin^2\vartheta P(\vartheta)\mathrm{d}\varphi^2\right),
\end{equation}
where
\begin{equation}
Q(r)=(1-\alpha^2r^2)\left(1-\frac{2m}{r}+\frac{e^2}{r^2}\right)-\frac{\Lambda}{3}r^2,
\end{equation}
\begin{equation}
P(\vartheta)=1-2\alpha m\cos\vartheta+\alpha^2e^2\cos^2\vartheta,
\end{equation}
\begin{equation}
\Omega(r,\vartheta)=1-\alpha r\cos\vartheta.
\end{equation}
When we set $\alpha=0$ the metric reduces to the Reissner-Nordstr\"{o}m-de Sitter
family of spacetimes including the Schwarzschild metric ($e=0$, $\Lambda=0$),
the Reissner-Nordstr\"{o}m metric ($\Lambda=0$), and the Schwarzschild-de Sitter metric
($e=0$). For $e=0$ and $\Lambda=0$ the metric reduces to the regular C-metric. For
$e=0$ it reduces to the C-de Sitter metric and for $\Lambda=0$ it reduces to the
charged C-metric. In accordance with observations we choose $0<m$. Theoretically
the electric charge $e$ can take any real value but because the line element (\ref{eq:line})
only contains its square we restrict it such that we have $0\leq e\leq e_{\mathrm{C}}$.
Because we live in an expanding Universe we will assume that $0\leq\Lambda<\Lambda_{\mathrm{C}}$.
$\alpha$ can be restricted from the symmetry of the spacetime. When we replace
$\alpha$ by $-\alpha$ and substitute $\vartheta \rightarrow \pi-\vartheta$ we see
that (\ref{eq:line}) is invariant and consequently we can restrict the
acceleration parameter to $0<\alpha<\alpha_{\mathrm{C}}$. Here, $e_{\mathrm{C}}$, $\Lambda_{\mathrm{C}}$ and
$\alpha_{\mathrm{C}}$ are critical parameters that have to be chosen according to the desired
interpretation of the spacetime.
In the charged C-de Sitter metric the time coordinate $t$ can assume any real value.
In addition we choose the angular coordinates $\vartheta$ and $\varphi$ such that
they represent the angular coordinates on the two sphere $S^2$. The range of the
$r$ coordinate is determined by the singularity structure of the spacetime, in
particular the singularities arising from $Q(r)$ and $\Omega(r,\vartheta)$. We will discuss
these singularities and their implications for the $r$ coordinate in the following.
The spacetime admits several singularities. We start by discussing the singularities
arising from $Q(r)$. In total $Q(r)$ can lead to up to five singularities in the metric.
The equation $Q(r)=0$ can lead to up to four singularities. In addition the metric
has another singularity at $r=0$. The singularity at $r=0$ is a curvature singularity.
For $e\neq0$ the curvature singularity is timelike. Because lightlike and timelike
geodesics cannot cross the curvature singularity particle motion is limited to
$0<r$. In addition the conformal factor becomes singular when $\Omega(r,\vartheta)=0$.
This singularity corresponds to conformal infinity and limits the radius coordinate
$r$. The singularity starts at $r=1/\alpha$ on the axis $\vartheta=0$ and extends
to $r=\infty$ for $\pi/2\leq\vartheta$. The other singularities are only coordinate
singularities $r_{\mathrm{H}}$ that can be removed using appropriate coordinate
transformations (for the C-metric the procedure is demonstrated in Ref. \citenum{Griffiths2006}).
In this article we restrict to black hole spacetimes and consequently all coordinate
singularities with $0<r_{\mathrm{H}}$ are horizons. In general the horizon struture
of each metric depends on the parameters of the spacetime. For each spacetime the
critical parameters $e_{\mathrm{C}}$, $\Lambda_{\mathrm{C}}$ and $\alpha_{\mathrm{C}}$
have to be chosen such that all horizons are real and that they lead to physically
meaningful black hole spacetimes. The horizon structure of the whole family of charged
C-de Sitter metrics is shown in Fig.~1. Panel a) shows the horizon struture of the
C-metric. The C-metric has two horizons, a black hole horizon at $r_{\mathrm{BH}}=2m$
and an acceleration horizon at $r_{\mathrm{BH}}<r_{\alpha}=1/\alpha$. Between the
horizons the spacetime is static. In the other two regions the spacetime is non-static.
Panels b) and c) show the horizon struture for the charged C-metric. The main difference
to the C-metric is that the spacetime now contains an inner and an outer black hole
horizon $r_{\mathrm{BH},\mathrm{i}}\leq r_{\mathrm{BH},\mathrm{o}}<r_{\alpha}$.
When $e=m$ both horizons coincide and form a degenerate horizon at $r_{\mathrm{BH}}=m$.
The regions $0<r<r_{\mathrm{BH},\mathrm{i}}$ and $r_{\mathrm{BH},\mathrm{o}}<r<r_{\alpha}$
are static. The other two regions are non-static. When we include the cosmological
constant $\Lambda$ (panels d) to f)) the radius coordinates of the horizons change,
however, the horizon structure remains unchanged. For a more detailed discussion
of the horizon structure in the C-metric we refer the interested reader to Ref.~\citenum{Frost2021}.
In our case we always choose $e_{\mathrm{C}}$ such that $P(\vartheta)$ does not lead to singularities in the metric.
Finally, the spacetime has two singularities on the axes at $\sin\vartheta=0$. As
demonstrated in Refs.~\citenum{Griffiths2009} and \citenum{Griffiths2006} these
singularities are conical singularites. They are associated with a deficit angle
($\vartheta=0$) and a surplus angle ($\vartheta=\pi$).
The charged C-de Sitter metric is usually interpreted such that it describes a
charged accelerating black hole with cosmological constant. In its full analytic
extension the spacetime describes two causally separated charged black holes accalerating
away from each other. The conical singularity at $\vartheta=0$ is commonly interpreted
as a string pulling the black hole, while the conical singularity at $\vartheta=\pi$
is interpreted as a strut pushing the black hole. As demonstrated in Ref.~\citenum{Griffiths2009} and \citenum{Griffiths2006}
one of the conical singularities can always be removed, however, as in our previous
work\cite{Frost2021} here we will retain both singularities to show the effects
of both, the string and the strut on geodesic motion.

\section{Equations of Motion}\label{sec:EoM}
In the charged C-de Sitter metric the equations of motion for light rays are fully
separable. In Mino parameterisation\cite{Mino2003} they read\cite{Grenzebach2015a,Frost2021}
\begin{equation}\label{eq:EoMt}
\frac{\mathrm{d}t}{\mathrm{d}\lambda}=\frac{r^2E}{Q(r)},
\end{equation}
\begin{equation}\label{eq:EoMr}
\left(\frac{\mathrm{d}r}{\mathrm{d}\lambda}\right)^2=E^2r^4-r^2Q(r)K,
\end{equation}
\begin{equation}\label{eq:EoMtheta}
\left(\frac{\mathrm{d}\vartheta}{\mathrm{d}\lambda}\right)^2=P(\vartheta)K-\frac{L_{z}^2}{\sin^2\vartheta},
\end{equation}
\begin{equation}\label{eq:EoMphi}
\frac{\mathrm{d}\varphi}{\mathrm{d}\lambda}=\frac{L_{z}}{\sin^2\vartheta P(\vartheta)},
\end{equation}
where the Mino parameter $\lambda$ is related to the affine parameter $s$ by
\begin{equation}\label{eq:Mino}
\frac{\mathrm{d}\lambda}{\mathrm{d}s}=\frac{\Omega(r,\vartheta)^2}{r^2}.
\end{equation}
Here, the three constants of motion $E$, $L_{z}$ and $K$ are the energy, the angular
momentum about the $z$ axis and the Carter constant of the light ray, respectively.
In this article we choose the energy $E$ such that $\mathrm{d}t/\mathrm{d}\lambda>0$.
In the following we will briefly discuss the equations of motion, the turning
points of the $r$ and the $\vartheta$ motion and identify the locations of the photon
sphere and the photon cone. Then we will turn to solving the equations of motion.
We will derive the solutions to the equations of motion for arbitrary initial
conditions $(x_{i}^{\mu})=(x^{\mu}(\lambda_{i}))=(t_{i},r_{i},\vartheta_{i},\varphi_{i})$
closely following the procedures described in Refs. \citenum{Frost2021} and \citenum{Gralla2020}.
Because the main focus of this article will be to apply our results to gravitational
lensing we will limit our discussion to lightlike geodesics in the region of outer
communication between photon sphere and the cosmological/acceleration horizon.

\subsection{The $r$ Motion}\label{sec:r}
\begin{figure}[ht]\label{fig:Vrpot}
  \centering
		\includegraphics[width=\textwidth]{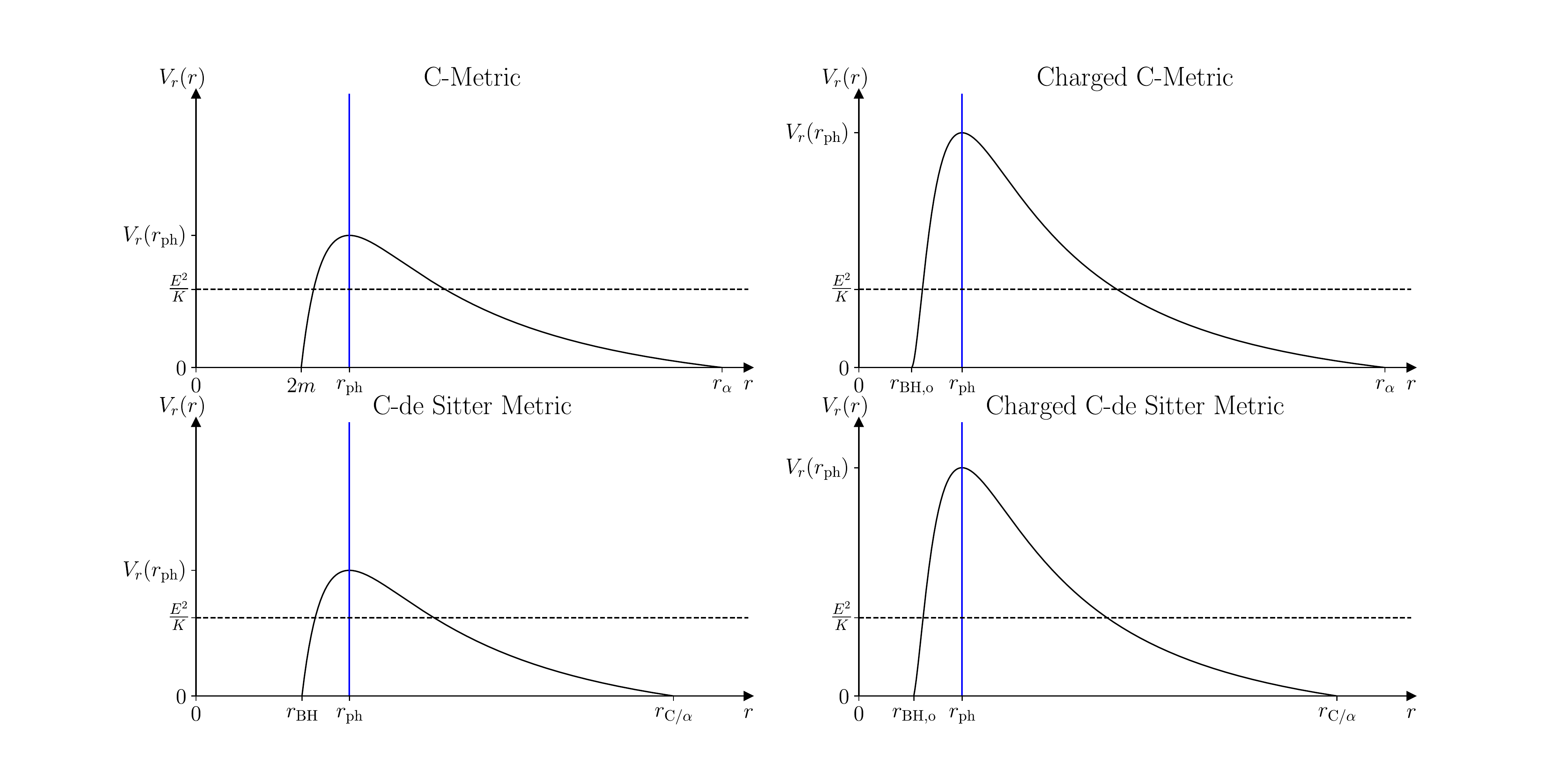}
	\caption{Potential $V_{r}(r)$ of the $r$ motion in the C-metric (top left), the charged C-metric (top right),
  the C-de Sitter metric (bottom left) and the charged C-de Sitter metric (bottom right) for $\Lambda=1/(200m^2)$, $e=m$,
  and $\alpha=1/(10m)$. The axes have the same scale in all four plots.}
\end{figure}
We begin with discussing the $r$ motion. Following Ref. \citenum{Frost2021} we
first rewrite (\ref{eq:EoMr}) in terms of the potential $V_{r}(r)$
\begin{equation}\label{eq:EoMrPot}
\frac{1}{r^4K}\left(\frac{\mathrm{d}r}{\mathrm{d}\lambda}\right)^2+V_{r}(r)=\frac{E^2}{K},
\end{equation}
where
\begin{equation}\label{eq:VrPot}
V_{r}(r)=\left(\frac{1}{r^2}-\alpha^2\right)\left(1-\frac{2m}{r}+\frac{e^2}{r^2}\right)-\frac{\Lambda}{3}.
\end{equation}
Fig.~2 shows the potentials for the C-metric (top left), the charged C-metric (top right),
the C-de Sitter metric (bottom left) and the charged C-de Sitter metric (bottom right). For all four metrics
the potential has a similar structure. The most striking difference occurs close
to the (outer) black hole horizon. For $e\neq0$ the maximum of $V_{r}(r)$ is much
higher than in the uncharged case. In addition for $\Lambda>0$ $V_{r}(r)$ approaches
zero at a smaller radius coordinate $r_{\mathrm{C}/\alpha}<r_{\alpha}$. The maximum of the
potential at $E^2/K=V_{r}(r_{\mathrm{ph}})$ marks the radius coordinate of an unstable photon
sphere. We can calculate the radius coordinate of the photon sphere by calculating
$\mathrm{d}r/\mathrm{d}\lambda=\mathrm{d}^2r/\mathrm{d}\lambda^2=0$. Combining both
conditions leads to the determining equation
\begin{eqnarray}\label{eq:rph}
r^3+\frac{1-\alpha^2e^2}{m\alpha^2}r^2-\frac{3}{\alpha^2}r+\frac{2e^2}{m\alpha^2}=0
\end{eqnarray}
for the radius of the photon sphere $r_{\mathrm{ph}}$. It is remarkable that although
the spherical symmetry is broken by the acceleration parameter the cosmological
constant does not have any effect on the radius coordinate of the photon sphere. The roots of this equation
can be analytically obtained using Cardano's method. As long as $e^2<1/\alpha^2$
we get three real solutions. Only the largest root lies in the region of outer communication
$r_{\mathrm{BH},\mathrm{o}}<r<r_{\mathrm{C}/\alpha}$ and thus marks the position of the photon sphere.
When $e=0$ the polynomial reduces to second order and $r_{\mathrm{ph}}$ is given by\cite{AlraisAlawadi2021,Frost2021}
\begin{equation}
  r_{\mathrm{ph}}=\frac{6m}{1+\sqrt{1+12\alpha^2m^2}}.
\end{equation}
We can read from Fig.~2 that at $r_{\mathrm{ph}}$ $V_{r}(r_{\mathrm{ph}})$ has always
a maximum and thus orbits of geodesics on the photon sphere are unstable. Here,
unstable means that if the orbit of a lightlike geodesic on the photon sphere
was radially infinitessimally perturbed the light ray would either fall into the
black hole or escape to (conformal) infinity.
Using the potential we can distinguish three different types of lightlike motion.
For $E^2/K>V_{r}(r_{\mathrm{ph}})$ lightlike geodesics have no turning points. Here we have
to distinguish between the cases $K=0$ (we will see in Section~\ref{sec:theta}
that these geodesics are principal null geodesics) and $K>0$. For $E^2/K>V_{r}(r_{\mathrm{ph}})$
and $K=0$ (\ref{eq:EoMr}) was already solved in Ref.~\citenum{Frost2021}. In
this case the solution reads:
\begin{equation}
r(\lambda)=\frac{r_{i}}{1-i_{r_{i}}r_{i}E\left(\lambda-\lambda_{i}\right)},
\end{equation}
where $i_{r_{i}}=\mathrm{sgn}\left(\left.\mathrm{d}r/\mathrm{d}\lambda\right|_{r=r_{i}}\right)$.
In the case $E^2/K>V_{r}(r_{\mathrm{ph}})$ and $K>0$ (\ref{eq:EoMr}) has two real and two
complex conjugated roots. We label them such that $r_{1}>r_{2}$ and $r_{3}=\bar{r}_{4}=R_{3}+iR_{4}$.
To solve (\ref{eq:EoMr}) we now substitute\cite{Gralla2020,Hancock1917}
\begin{equation}
r=\frac{r_{1}\bar{R}-r_{2}R+(r_{1}\bar{R}+r_{2}R)\cos\chi_{r}}{\bar{R}-R+(\bar{R}+R)\cos\chi_{r}},
\end{equation}
where
\begin{equation}\label{eq:RbarR}
R=\sqrt{(R_{3}-r_{1})^2+R_{4}^2}~~~\mathrm{and}~~~\bar{R}=\sqrt{(R_{3}-r_{2})^2+R_{4}^2}.
\end{equation}
Then we follow the steps described in Appendix B of Ref.~\citenum{Frost2021} and
obtain the solution in terms of Jacobi's elliptic $\mathrm{cn}$ function. It reads
\begin{equation}
r(\lambda)=\frac{r_{1}\bar{R}-r_{2}R+(r_{1}\bar{R}+r_{2}R)\mathrm{cn}\left(a_{r}\left(\lambda-\lambda_{i}\right)+\lambda_{r_{i},k_{1}},k_{1}\right)}{\bar{R}-R+(\bar{R}+R)\mathrm{cn}\left(a_{r}\left(\lambda-\lambda_{i}\right)+\lambda_{r_{i},k_{1}},k_{1}\right)},
\end{equation}
where
\begin{eqnarray}
a_{r}=i_{r_{i}}\sqrt{\left(E^2+\left(\alpha^2+\frac{\Lambda}{3}\right)K\right)R\bar{R}},& \lambda_{r_{i},k_{1}}=F_{L}(\chi_{r_{i}},k_{1}), \\
\chi_{r_{i}}=\mathrm{arccos}\left(\frac{(r_{i}-r_{2})R-(r_{i}-r_{1})\bar{R}}{(r_{i}-r_{2})R+(r_{i}-r_{1})\bar{R}}\right),&   k_{1}=\frac{(R+\bar{R})^2-(r_{1}-r_{2})^2}{4R\bar{R}}.\nonumber
\end{eqnarray}
The condition $E^2/K=V_{r}(r_{\mathrm{ph}})$ characterises lightlike geodesics asymptotically
coming from or going to the photon sphere. In this case (\ref{eq:EoMr}) has four
real roots, two of which are equal. We label them such that $r_{1}=r_{2}=r_{\mathrm{ph}}>r_{3}>r_{4}$.
In this case we substitute
\begin{equation}
r=r_{3}+\frac{3a_{3,r}}{12y-a_{2,r}},
\end{equation}
where
\begin{equation}
a_{2,r}=6\left(E^2+\left(\alpha^2+\frac{\Lambda}{3}\right)K\right)r_{3}^2-6m\alpha^2 K r_{3}-(1-\alpha^2e^2)K,
\end{equation}
\begin{equation}
a_{3,r}=4\left(E^2+\left(\alpha^2+\frac{\Lambda}{3}\right)K\right)r_{3}^3-6m\alpha^2 K r_{3}^2-2(1-\alpha^2e^2)Kr_{3}+2mK.
\end{equation}
Then we follow the steps described in Section 3.2.3 in Ref.~\citenum{Frost2021}
and obtain as solution for $r(\lambda)$:
\begin{equation}
r(\lambda)=r_{3}-\frac{(r_{\mathrm{ph}}-r_{3})(r_{3}-r_{4})}{r_{\mathrm{ph}}-r_{3}-(r_{\mathrm{ph}}-r_{4})\tanh^2\left(b_{r}+i_{r_{i}}\sqrt{\frac{a_{3,r}(r_{\mathrm{ph}}-r_{4})}{4(r_{\mathrm{ph}}-r_{3})(r_{3}-r_{4})}}(\lambda_{i}-\lambda)\right)},
\end{equation}
where
\begin{equation}
b_{r}=\mathrm{artanh}\left(\sqrt{\frac{(r_{i}-r_{4})(r_{\mathrm{ph}}-r_{3})}{(r_{i}-r_{3})(r_{\mathrm{ph}}-r_{4})}}\right).
\end{equation}
All remaining lightlike geodesics have $E^2/K<V_{r}(r_{\mathrm{ph}})$. These geodesics have
four real roots and can pass through a turning point. Outside the photon sphere
this turning point is always a minimum. For solving (\ref{eq:EoMr}) we now label
the roots such that $r_{1}=r_{\mathrm{min}}>r_{2}>r_{3}>r_{4}$. Then we substitute\cite{Gralla2020,Hancock1917}
\begin{equation}
r=r_{2}+\frac{(r_{1}-r_{2})(r_{2}-r_{4})}{r_{2}-r_{4}-(r_{1}-r_{4})\sin^2\chi_{r}}.
\end{equation}
Again we apply the steps described in Appendix B of Ref.~\citenum{Frost2021}. This
time we obtain the solution to (\ref{eq:EoMr}) in terms of Jacobi's elliptic $\mathrm{sn}$
function
\begin{equation}
r(\lambda)=r_{2}+\frac{(r_{1}-r_{2})(r_{2}-r_{4})}{r_{2}-r_{4}-(r_{1}-r_{4})\mathrm{sn}^2\left(c_{r}\left(\lambda-\lambda_{i}\right)+\lambda_{r_{i},k_{2}},k_{2}\right)},
\end{equation}
where
\begin{eqnarray}
c_{r}=\frac{i_{r_{i}}}{2}\sqrt{\left(E^2+\left(\alpha^2+\frac{\Lambda}{3}\right)K\right)(r_{1}-r_{3})(r_{2}-r_{4})},& \lambda_{r_{i},k_{2}}=F_{L}(\chi_{r_{i}},k_{2}), \\
\chi_{r_{i}}=\mathrm{arcsin}\left(\sqrt{\frac{(r_{i}-r_{1})(r_{2}-r_{4})}{(r_{i}-r_{2})(r_{1}-r_{4})}}\right),&   k_{2}=\frac{(r_{2}-r_{3})(r_{1}-r_{4})}{(r_{1}-r_{3})(r_{2}-r_{4})}.\nonumber
\end{eqnarray}

\subsection{The $t$ Coordinate}\label{eq:solt}
Now we calculate the time coordinate $t$. The right-hand side of (\ref{eq:EoMt})
depends only on $r(\lambda)$. Therefore, we first rewrite (\ref{eq:EoMt}) as a
differential of $r$. For this purpose we divide (\ref{eq:EoMt}) by the root of
(\ref{eq:EoMr}). Then we integrate over $r$ starting at $t(\lambda_{i})=t(r_{i})=t_{i}$.
The resulting integral reads
\begin{equation}\label{eq:Intt}
t(\lambda)=t_{i}+\int_{r_{i}...}^{...r(\lambda)}\frac{E r'^2\mathrm{d}r'}{Q(r')\sqrt{E^2r'^4-r'^2Q(r')K}}.
\end{equation}
Here, we have to choose the sign of the root in accordance with the $r$ motion and
the dots in the limits of the integration indicate that we have to split the integral
at the turning points. The procedure to integrate (\ref{eq:Intt}) is straight forward. We have to distinguish
the same types of motion as for $r(\lambda)$ in Section~\ref{sec:r}. First
we perform a partial fraction decomposition of $r'^2/Q(r')$. Then we follow
the procedure described in Refs.~\citenum{Frost2021} and \citenum{Gralla2020} to express (\ref{eq:Intt})
in terms of elementary functions and elliptic integrals. Note that some of the elliptic
integrals cannot immediately be expressed in form of elementary functions and elliptic
integrals of first, second and third kind. However, we can easily rewrite them using
the procedures described in Appendix B in Ref.~\citenum{Gralla2020} and Appendix A
in Ref.~\citenum{Frost2021}.

\subsection{The $\vartheta$ Motion}\label{sec:theta}
\begin{figure}[ht]\label{fig:Vthetapot}
  \centering
		\includegraphics[width=\textwidth]{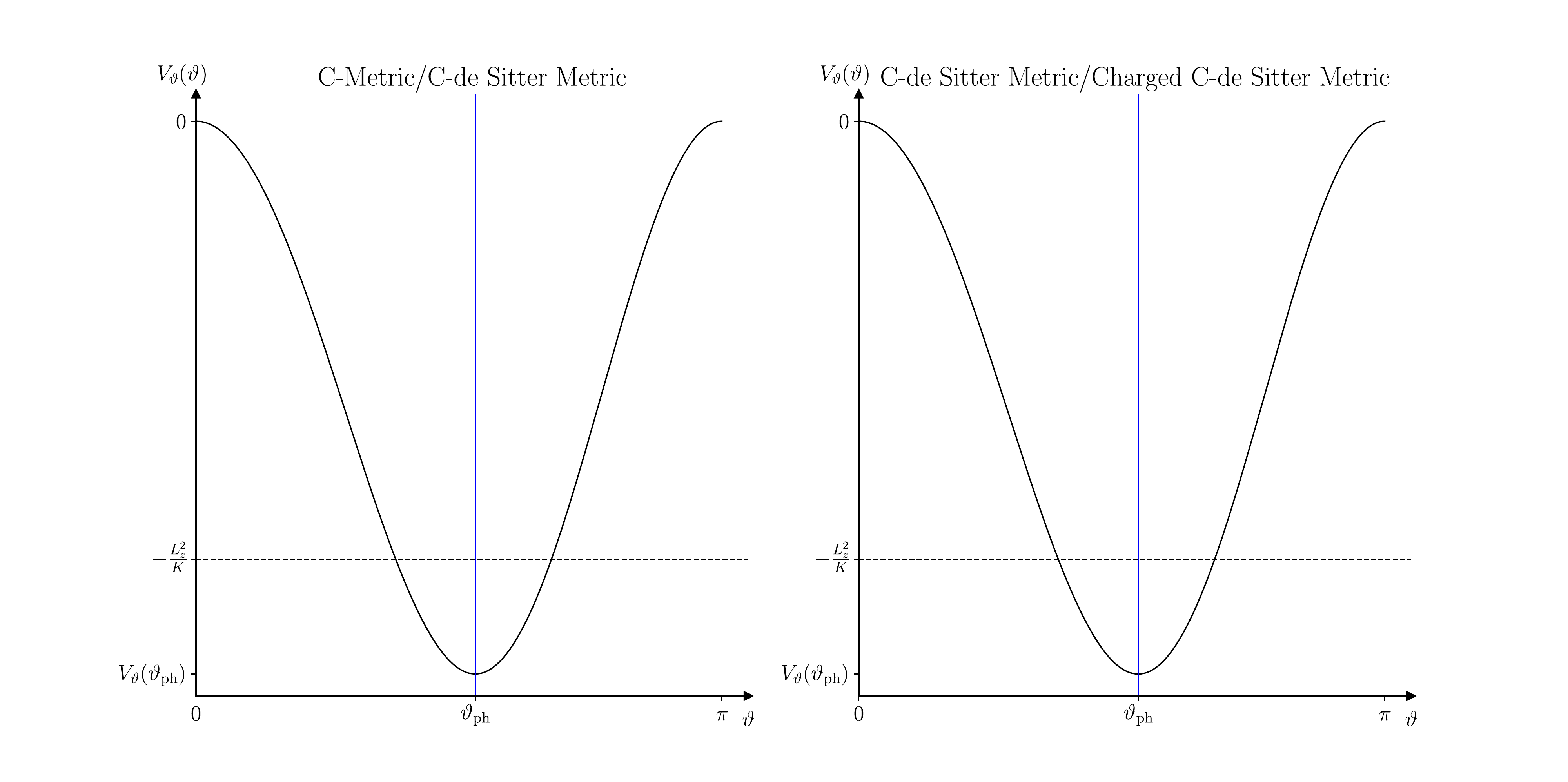}
	\caption{Potential $V_{\vartheta}(\vartheta)$ of the $\vartheta$ motion in the
  C-metric/C-de Sitter metric (left) and the charged C-metric/C-de Sitter metric (right) for $e=m$
  and $\alpha=1/(10m)$. The axes have the same scale in both plots.}
\end{figure}
Again we start by rewriting (\ref{eq:EoMtheta}) in terms of the potential $V_{\vartheta}(\vartheta)$
\begin{equation}
\frac{\sin^2\vartheta}{K}\left(\frac{\mathrm{d}\vartheta}{\mathrm{d}\lambda}\right)^2+V_{\vartheta}(\vartheta)=-\frac{L_{z}^2}{K},
\end{equation}
where
\begin{equation}
V_{\vartheta}(\vartheta)=-\sin^2\vartheta(1-2\alpha m\cos\vartheta+\alpha^2e^2\cos^2\vartheta).
\end{equation}
Fig.~3 shows the potentials $V_{\vartheta}(\vartheta)$ of the
$\vartheta$ motion for the C-(de Sitter) metric (left) and the charged C-(de Sitter) metric (right).
Although in the right plot we set $e=m$ the differences are invisibly small. The
potential has always a minimum at a single value $\vartheta$. At this minimum we
have $\mathrm{d}\vartheta/\mathrm{d}\lambda=\mathrm{d}^2\vartheta/\mathrm{d}\lambda^2=0$.
Combining both conditions leads to the determining equation
\begin{eqnarray}\label{eq:thetaph}
\cos^3\vartheta-\frac{3m}{2\alpha e^2}\cos^2\vartheta+\frac{1-\alpha^2e^2}{2\alpha^2e^2}\cos\vartheta+\frac{m}{2\alpha e^2}=0.
\end{eqnarray}
We determine the roots of this equation using Cardano's method. Within the permissible
range $0\leq\vartheta\leq\pi$ the equation has exactly one real root $\vartheta_{\mathrm{ph}}$.
This is the photon cone of the charged C-(de Sitter) metric. Note that we cannot
determine the limits $e\rightarrow0$ or $\alpha\rightarrow 0$ from the calculated
$\vartheta_{\mathrm{ph}}$ directly. When we approach these limits we see that for $e\rightarrow 0$
$\vartheta_{\mathrm{ph}}$ reduces to\cite{AlraisAlawadi2021,Frost2021}
\begin{equation}
\vartheta_{\mathrm{ph}}=\mathrm{arccos}\left(\frac{-2\alpha m}{1+\sqrt{1+12\alpha^2 m^2}}\right)
\end{equation}
and for $\alpha\rightarrow0$ it reduces to $\vartheta_{\mathrm{ph}}=\pi/2$. The photon cone
is unstable. Here unstable means that when light rays on the photon cone are infinitessimally
perturbed in $\vartheta$ direction they will begin to oscillate between a minimum
$\vartheta_{\mathrm{min}}$ and a maximum $\vartheta_{\mathrm{max}}$.
Based on the potential $V_{\vartheta}(\vartheta)$ we can distinguish two different
types of motion. From (\ref{eq:EoMtheta}) we can immediately read that we have
$L_{z}=0$ when $K=0$ and the right-hand side of (\ref{eq:EoMtheta}) vanishes.
Because $L_{z}=0$ implies $\mathrm{d}\varphi/\mathrm{d}\lambda=0$ these are the
already in Section~\ref{sec:r} mentioned principal null geodesics.
Also for lightlike geodesics on the photon cone ($-L_{z}^2/K=V_{\vartheta}(\vartheta_{\mathrm{ph}})$)
the right-hand side of (\ref{eq:EoMtheta}) vanishes because we have two coinciding
roots at $x=\cos\vartheta_{\mathrm{ph}}$. Thus for principal null geodesics and lightlike geodesics
on the photon cone the solution to (\ref{eq:EoMtheta}) reads $\vartheta(\lambda)=\vartheta_{i}$.
For all other lightlike geodesics we have $-L_{z}^2/K\neq V_{\vartheta}(\vartheta_{\mathrm{ph}})$.
These geodesics oscillate between a minimum $\vartheta_{\mathrm{min}}$ and a maximum $\vartheta_{\mathrm{max}}$
and each geodesic can potentially have arbitrary many turning points. In this case
we first rewrite (\ref{eq:EoMtheta}) in terms of $x=\cos\vartheta$. It is easy
to see that in the new parameterisation (\ref{eq:EoMtheta}) reduces to a polynomial
of third order in $x$ when $e=0$. This case was already treated in Ref.~\citenum{Frost2021}
and thus we do not reproduce it here. For $e\neq0$ we always have two turning points,
however, depending on the choice of $L_{z}$ and $K$ the right-hand side of (\ref{eq:EoMtheta})
has two different root structures. In the first case two of the roots are real.
We label them such that $x_{2}=\cos\vartheta_{\mathrm{max}}<x_{1}=\cos\vartheta_{\mathrm{min}}$.
The other two roots are complex conjugated and we label them as $x_{3}=\bar{x}_{4}=X_{3}+iX_{4}$.
As first step to solve (\ref{eq:EoMtheta}) we now substitute\cite{Hancock1917}
\begin{equation}\label{eq:subtheta1}
x=\frac{x_{1}\bar{X}+x_{2}X+(x_{1}\bar{X}-x_{2}X)\cos\chi_{\vartheta}}{\bar{X}+X+(\bar{X}-X)\cos\chi_{\vartheta}},
\end{equation}
where $X$ and $\bar{X}$ are given by (\ref{eq:RbarR}). In the second step we again
follow the steps described in Appendix B of Ref.~\citenum{Frost2021}. We obtain
$\vartheta(\lambda)$ in terms of Jacobi's elliptic $\mathrm{cn}$ function. It reads
\begin{equation}
\vartheta(\lambda)=\mathrm{arccos}\left(\frac{x_{1}\bar{X}+x_{2}X+(x_{1}\bar{X}-x_{2}X)\mathrm{cn}\left(a_{\vartheta}\left(\lambda-\lambda_{i}\right)+\lambda_{\vartheta_{i},k_{3}},k_{3}\right)}{\bar{X}+X+(\bar{X}-X)\mathrm{cn}\left(a_{\vartheta}\left(\lambda-\lambda_{i}\right)+\lambda_{\vartheta_{i},k_{3}},k_{3}\right)}\right),
\end{equation}
where $i_{\vartheta_{i}}=\mathrm{sgn}\left(\left.\mathrm{d}\vartheta/\mathrm{d}\lambda\right|_{\vartheta=\vartheta_{i}}\right)$ and
\begin{eqnarray}
a_{\vartheta}=i_{\vartheta_{i}}\sqrt{\alpha^2 e^2K X\bar{X}},& \lambda_{\vartheta_{i},k_{3}}=F_{L}(\chi_{\vartheta_{i}},k_{3}), \\
\chi_{\vartheta_{i}}=\mathrm{arccos}\left(\frac{(\cos\vartheta_{i}-x_{1})\bar{X}+(\cos\vartheta_{i}-x_{2})X}{(x_{1}-\cos\vartheta_{i})\bar{X}+(\cos\vartheta_{i}-x_{2})X}\right),& k_{3}=\frac{(x_{1}-x_{2})^2-(X-\bar{X})^2}{4X\bar{X}}.\nonumber
\end{eqnarray}
In the second case all four roots are real and we label them such that $x_{4}=\cos\vartheta_{\mathrm{max}}<x_{3}=\cos\vartheta_{\mathrm{min}}<x_{2}<x_{1}$.
In this case we substitute\cite{Hancock1917}
\begin{equation}\label{eq:subtheta2}
x=x_{1}-\frac{(x_{1}-x_{3})(x_{1}-x_{4})}{x_{1}-x_{3}+(x_{3}-x_{4})\sin^2\chi_{\vartheta}}.
\end{equation}
We again follow the steps described in Appendix B of Ref.~\citenum{Frost2021}. This
time we obtain $\vartheta(\lambda)$ in terms of Jacobi's elliptic $\mathrm{sn}$ function
\begin{equation}
\vartheta(\lambda)=\mathrm{arccos}\left(x_{1}-\frac{(x_{1}-x_{3})(x_{1}-x_{4})}{x_{1}-x_{3}+(x_{3}-x_{4})\mathrm{sn}^2\left(b_{\vartheta}\left(\lambda_{i}-\lambda\right)+\lambda_{\vartheta_{i},k_{4}},k_{4}\right)}\right),
\end{equation}
where
\begin{eqnarray}
b_{\vartheta}=\frac{i_{\vartheta_{i}}}{2}\sqrt{\alpha^2 e^2K(x_{1}-x_{3})(x_{2}-x_{4})},& \lambda_{\vartheta_{i},k_{4}}=F_{L}(\chi_{\vartheta_{i}},k_{4}) \\
\chi_{\vartheta_{i}}=\mathrm{arcsin}\left(\sqrt{\frac{(\cos\vartheta_{i}-x_{4})(x_{1}-x_{3})}{(x_{1}-\cos\vartheta_{i})(x_{3}-x_{4})}}\right),& k_{4}=\frac{(x_{1}-x_{2})(x_{3}-x_{4})}{(x_{1}-x_{3})(x_{2}-x_{4})}.\nonumber
\end{eqnarray}

\subsection{The $\varphi$ Motion}\label{sec:phi}
The $\varphi$ motion is governed by (\ref{eq:EoMphi}). We immediately read that
for $L_{z}=0$ the right-hand side vanishes. For $K=0$ these are the principal null
geodesics. In this case we obtain the solution to (\ref{eq:EoMphi}) as $\varphi(\lambda)=\varphi_{i}$.
If $K\neq0$ these are geodesics crossing the axes. What happens to these geodesics
now depends on whether we assume that the string and the strut are opaque or transparent.
In the former case the lightlike geodesics terminate at the string and we observe
it as a black line blocking out all light emitted by sources located behind it.
In the latter case the lightlike geodesics pass through the string. To see what
happens when a lightlike geodesic passes through the axes we consider a series of
lightlike geodesics continually approaching the axes for $L_{z}>0$ and $L_{z}<0$.\cite{Frost2021}
We observe that for $L_{z}>0$ and $L_{z}<0$ the $\varphi$ coordinate has different
limits. Hence geodesics crossing the axes have two different continuations and the
$\varphi$ coordinate is not continuous.\\
Next we turn to lightlike geodesics with a double root at $x=\cos\vartheta_{\mathrm{ph}}$ ($-L_{z}^2/K= V_{\vartheta}(\vartheta_{\mathrm{ph}})$).
These are lightlike geodesics on the photon cone with $\vartheta(\lambda)=\vartheta_{\mathrm{ph}}$.
In this case the right-hand side of (\ref{eq:EoMphi}) is constant and the solution
$\varphi(\lambda)$ reads
\begin{equation}\label{eq:solphiph}
\varphi(\lambda)=\varphi_{i}+\frac{L_{z}(\lambda-\lambda_{i})}{\sin^2\vartheta_{\mathrm{ph}}P(\vartheta_{\mathrm{ph}})}.
\end{equation}
Last we turn to lightlike geodesics with $-L_{z}^2/K\neq V_{\vartheta}(\vartheta_{\mathrm{ph}})$.
In this case we want to express $\varphi(\lambda)$ in terms of elementary functions
and elliptic integrals. For this purpose we first substitute $x=\cos\vartheta$ in
(\ref{eq:EoMtheta}) and (\ref{eq:EoMphi}). Then we divide (\ref{eq:EoMphi}) by
the root of (\ref{eq:EoMtheta}) and integrate. Now $\varphi(\lambda)$ reads
\begin{equation}\label{eq:solphi}
\varphi(\lambda)=\varphi_{i}+\int_{\cos\vartheta_{i}...}^{...\cos\vartheta(\lambda)}\frac{L_{z} \mathrm{d}x'}{(1-x'^2)P(x')\sqrt{(1-x'^2)P(x')K-L_{z}^2}}.
\end{equation}
Here, we have to choose the sign of the root such that it corresponds to the sign
of the $\cos\vartheta$ motion and the dots in the limits indicate that we have to
split the integral at the turning points. Again we rewrite the elliptic integral
using elementary functions and the canonical forms of the elliptic integrals. For
this purpose we now perform a partial fracion decomposition of $(1-x^2)^{-1}P(x)^{-1}$.
Then we use (41) in Ref.~\citenum{Frost2021} and (\ref{eq:subtheta1}) and
(\ref{eq:subtheta2}) to rewrite (\ref{eq:solphi}) in terms of elementary functions
and elliptic integrals of first, second and third kind. Note that we again encounter
elliptic integrals which do not immediately have a canonical form. Again we rewrite
them using the procedures described in Appendix B in Ref.~\citenum{Gralla2020} and
Appendix A in Ref.~\citenum{Frost2021} (although here we have $n^2/(n^2-1)<1$ and
not $1<n^2/(n^2-1)$ the basic integration procedure is the same).

\section{Gravitational Lensing}\label{sec:Lensing}
\subsection{Celestial Sphere}
In astronomy the position of a light source on the sky is identified using latitude
and longitude coordinates. For this purpose astronomers fix the main target of their
observation at the centre of their image and measure the position of all other objects
relative to the centre. In our discussion of gravitational lensing in the charged
C-de Sitter metric we follow this approach. For this purpose we first fix an observer
at coordinates $(x_{O}^{\mu})=(t_{O},r_{O},\vartheta_{O},\varphi_{O})$ in the region
of outer communication between photon sphere and cosmological/acceleration horizon.
Then we choose the black hole as main target of our observation and introduce an
orthonormal tetrad following Refs.~\citenum{Grenzebach2015a} and \citenum{Frost2021}:
\begin{eqnarray}
 e_{0}=\left.\frac{\Omega(r,\vartheta)}{\sqrt{Q(r)}}\partial_{t}\right|_{(x_{O}^{\mu})},~~~~e_{1}=\left.\frac{\Omega(r,\vartheta)\sqrt{P(\vartheta)}}{r}\partial_{\vartheta}\right|_{(x_{O}^{\mu})},
\end{eqnarray}
\begin{eqnarray}
 e_{2}=-\left.\frac{\Omega(r,\vartheta)}{r\sin\vartheta\sqrt{P(\vartheta)}}\partial_{\varphi}\right|_{(x_{O}^{\mu})},~~~~e_{3}=-\left.\Omega(r,\vartheta)\sqrt{Q(r)}\partial_{r}\right|_{(x_{O}^{\mu})}.\nonumber
\end{eqnarray}
We will call the angles on the celestial sphere surrounding the observer $\Sigma$
(latitude) and $\Psi$ (longitude). The angle $\Sigma$ is measured from the axis
along $e_{3}$ connecting lens and observer and the angle $\Psi$ is measured in the
direction of $e_{2}$ from the axis along the direction of $e_{1}$. \\
Before we can investigate gravitational lensing in the charged C-de Sitter metric
we first have to derive the relations between the constants of motion $E$, $L_{z}$
and $K$ and the angles on the observer's celestial sphere $\Sigma$ and $\Psi$.
For this purpose we consider a lightlike geodesic ending at the position
of the observer. Now we first write down the tangent vector to this geodesic. It
reads
\begin{eqnarray}\label{eq:lighttang}
 \frac{\mathrm{d}\eta}{\mathrm{d}\lambda}=\frac{\mathrm{d}t}{\mathrm{d}\lambda}\partial_{t}+\frac{\mathrm{d}r}{\mathrm{d}\lambda}\partial_{r}+\frac{\mathrm{d}\vartheta}{\mathrm{d}\lambda}\partial_{\vartheta}+\frac{\mathrm{d}\varphi}{\mathrm{d}\lambda}\partial_{\varphi}.
\end{eqnarray}
At the position of the observer we can also express the tangent vector of the geodesic
by the angles on the observer's celestial sphere and the tetrad vectors $e_{0}$,
$e_{1}$, $e_{2}$ and $e_{3}$
\begin{eqnarray}\label{eq:lighttetrad}
 \frac{\mathrm{d}\eta}{\mathrm{d}\lambda}=\sigma\left(-e_{0}+\sin\Sigma\cos\Psi e_{1}+\sin\Sigma\sin\Psi e_{2}+\cos\Sigma e_{3}\right),
\end{eqnarray}
where $\sigma$ is a normalisation constant
\begin{eqnarray}\label{eq:normC}
 \sigma=g\left(\frac{\mathrm{d}\eta}{\mathrm{d}\lambda},e_{0}\right).
\end{eqnarray}
In our convention we have $E>0$ and thus $\sigma$ has to be negative. Because the
Mino parameter $\lambda$ is defined up to an affine transformation without loss of
generality we can choose $\sigma=-r_{O}^2/\Omega(r_{O},\vartheta_{O})^2$.\cite{Frost2021} Now we insert
$\sigma$ in (\ref{eq:lighttetrad}) and (\ref{eq:normC}). Then a comparison of the
coefficients of (\ref{eq:lighttang}) and (\ref{eq:lighttetrad}) leads to\cite{Frost2021}
\begin{eqnarray}\label{eq:CoM}
E=\frac{\sqrt{Q(r_{O})}}{\Omega(r_{O},\vartheta_{O})},~L_{z}=\frac{r_{O}\sqrt{P(\vartheta_{O})}\sin\vartheta_{O}\sin\Sigma\sin\Psi}{\Omega(r_{O},\vartheta_{O})},~K=\frac{r_{O}^2\sin^2\Sigma}{\Omega(r_{O},\vartheta_{O})^2}.
\end{eqnarray}

\subsection{Angular Radius of the Photon Sphere}
The shadow of a black hole is a higly idealised concept. It is constructed as follows.
First we place an observer in the region of outer communication between photon
sphere and cosmological/acceleration horizon. Then we distribute
light sources everywhere except between observer and the black hole. Therefore
the former region is associated with brightness on the observer's sky while the
latter is associated with darkness. Now we shoot back light rays exactly on the boundary
between these areas of brightness and darkness. These are geodesics asymptotically
going to the photon sphere. They have the same constants of motion as light rays
on the photon sphere and thus a double root at $r_{\mathrm{ph}}$. We now use this fact to
calculate the angular radius of the shadow. For this purpose we first insert (\ref{eq:CoM})
in (\ref{eq:EoMr}). Then we employ that $\left.\mathrm{d}r/\mathrm{d}\lambda\right|_{r=r_{\mathrm{ph}}}=0$.
Now we solve for $\Sigma$ and get as angular radius of the shadow\cite{Frost2021}
\begin{equation}
\Sigma_{\mathrm{ph}}=\mathrm{arcsin}\left(\frac{r_{\mathrm{ph}}}{r_{O}}\sqrt{\frac{Q(r_{O})}{Q(r_{\mathrm{ph}})}}\right).
\end{equation}

\subsection{Lens Map}\label{sec:lensmap}
\begin{figure}\label{fig:LensEq}
\begin{tabular}{cc}
  C-Metric & Charged C-Metric \\[6pt]
  \includegraphics[width=65mm]{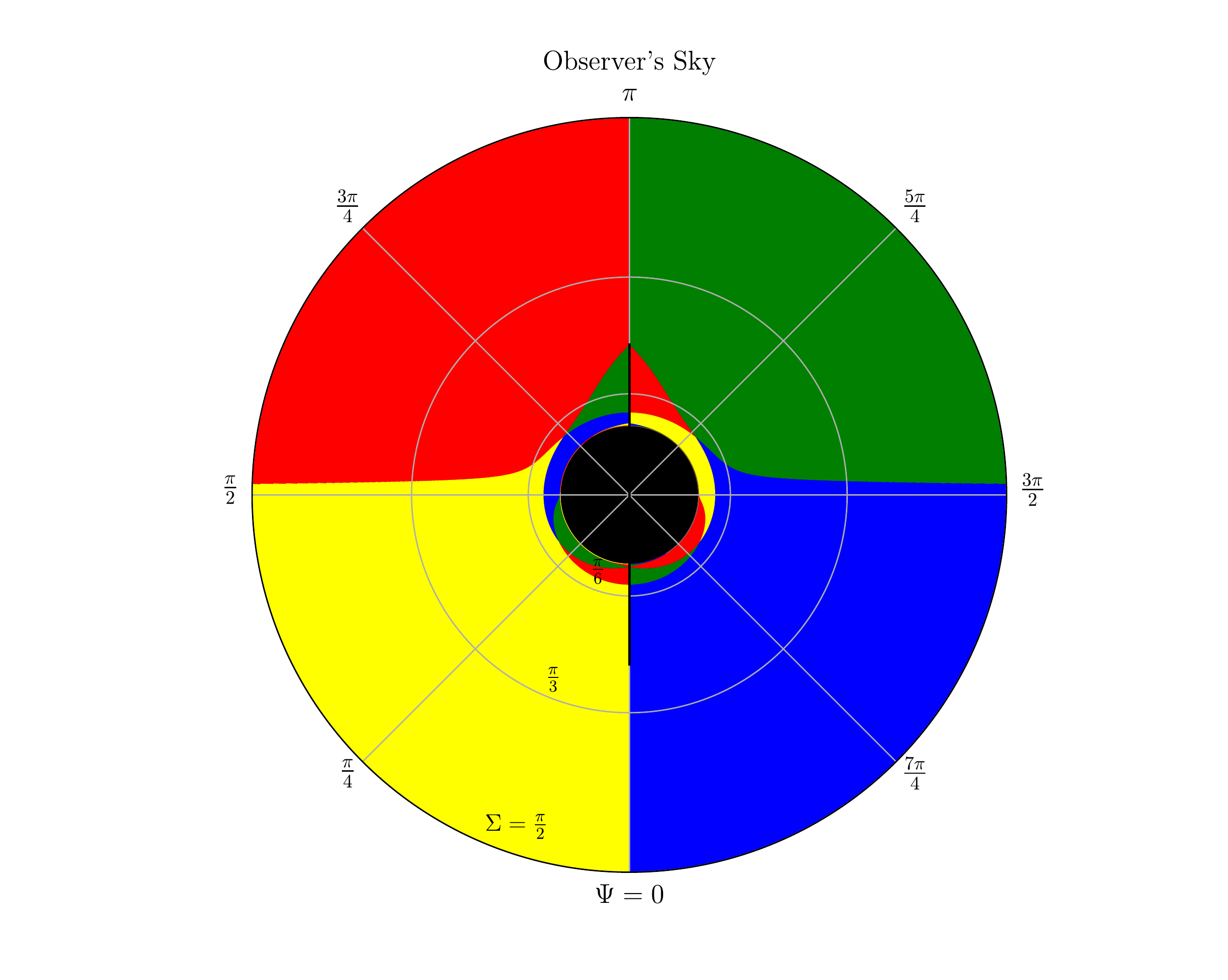} &   \includegraphics[width=65mm]{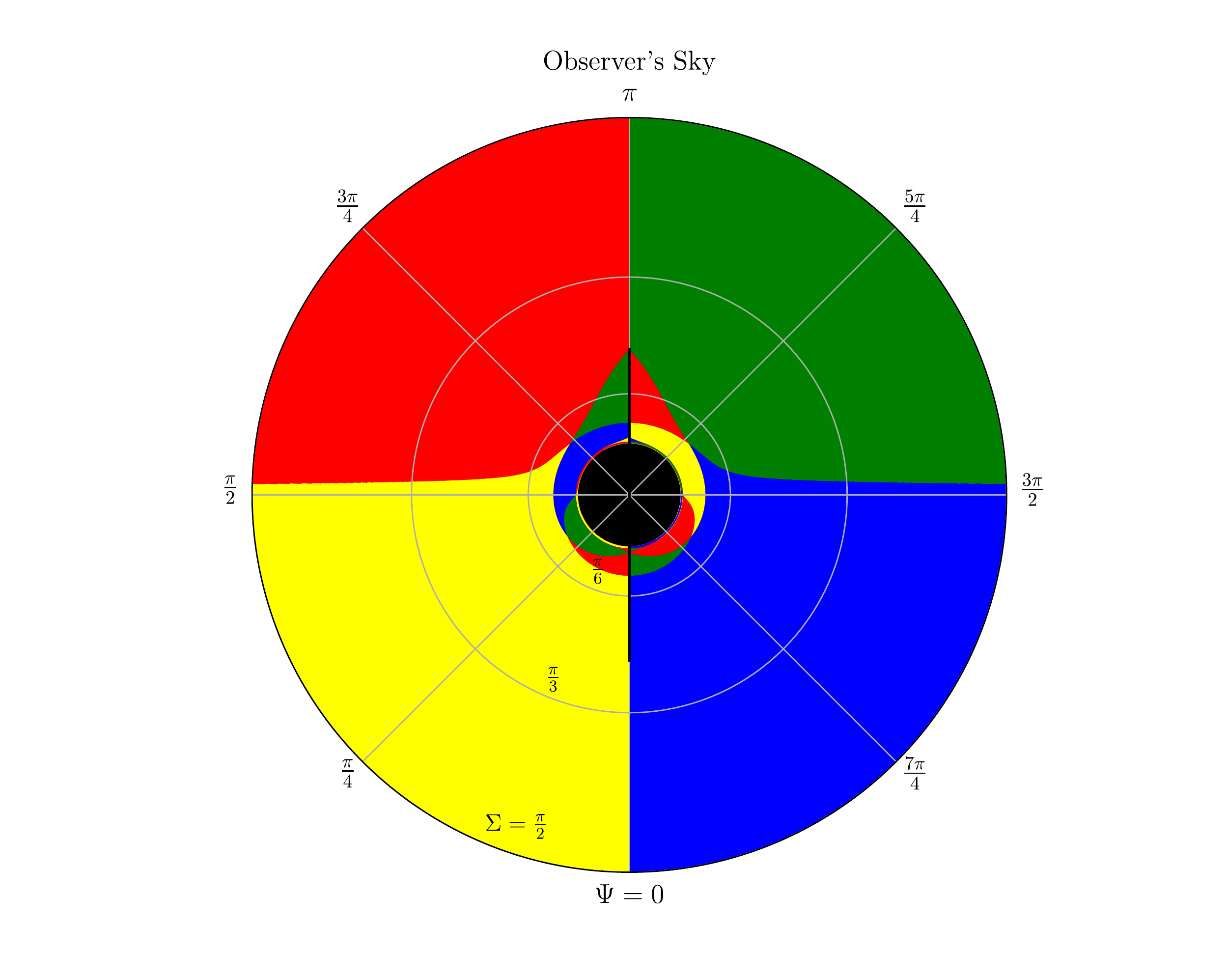} \\
  C-de Sitter Metric & Charged C-de Sitter Metric \\[6pt]
  \includegraphics[width=65mm]{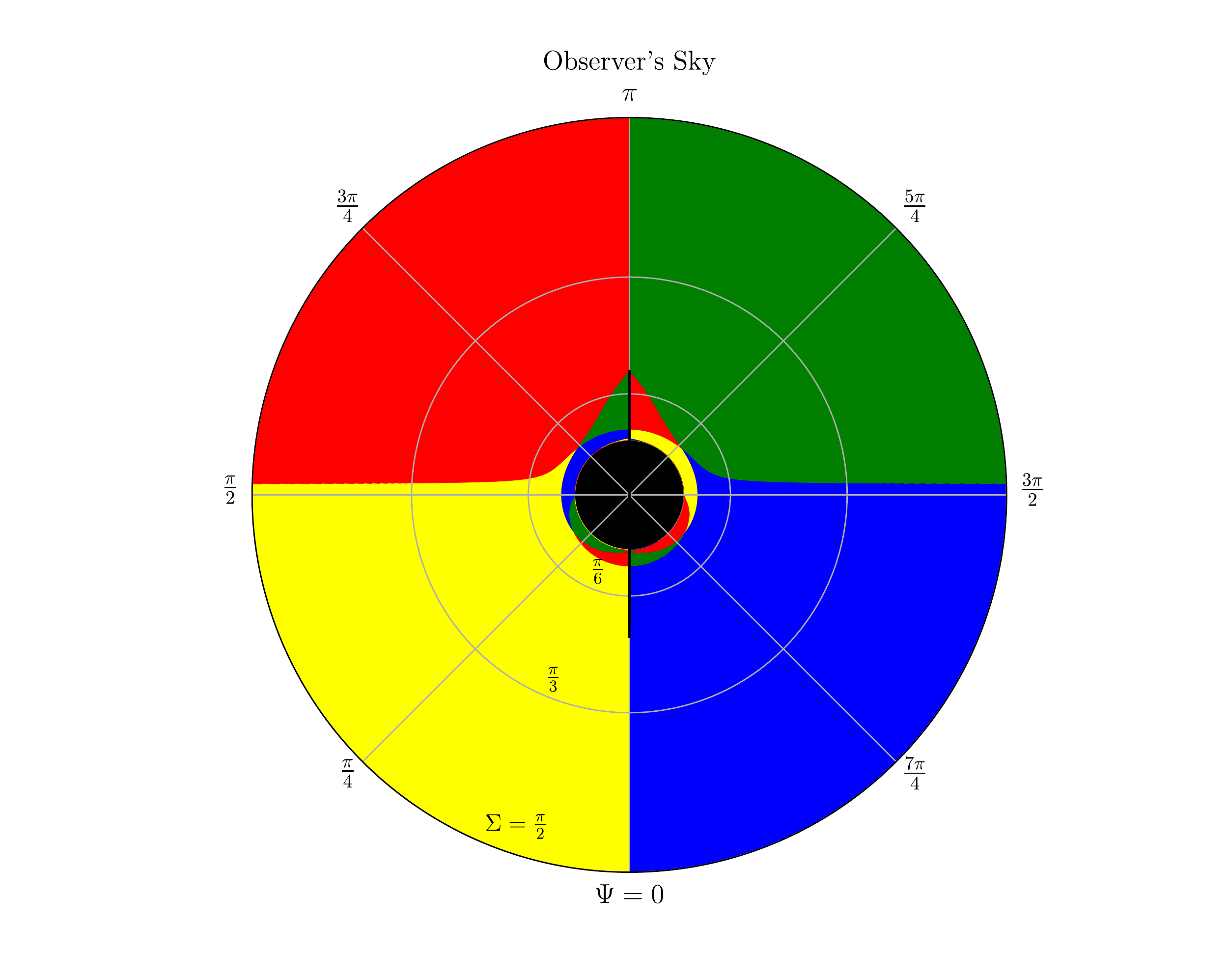} &   \includegraphics[width=65mm]{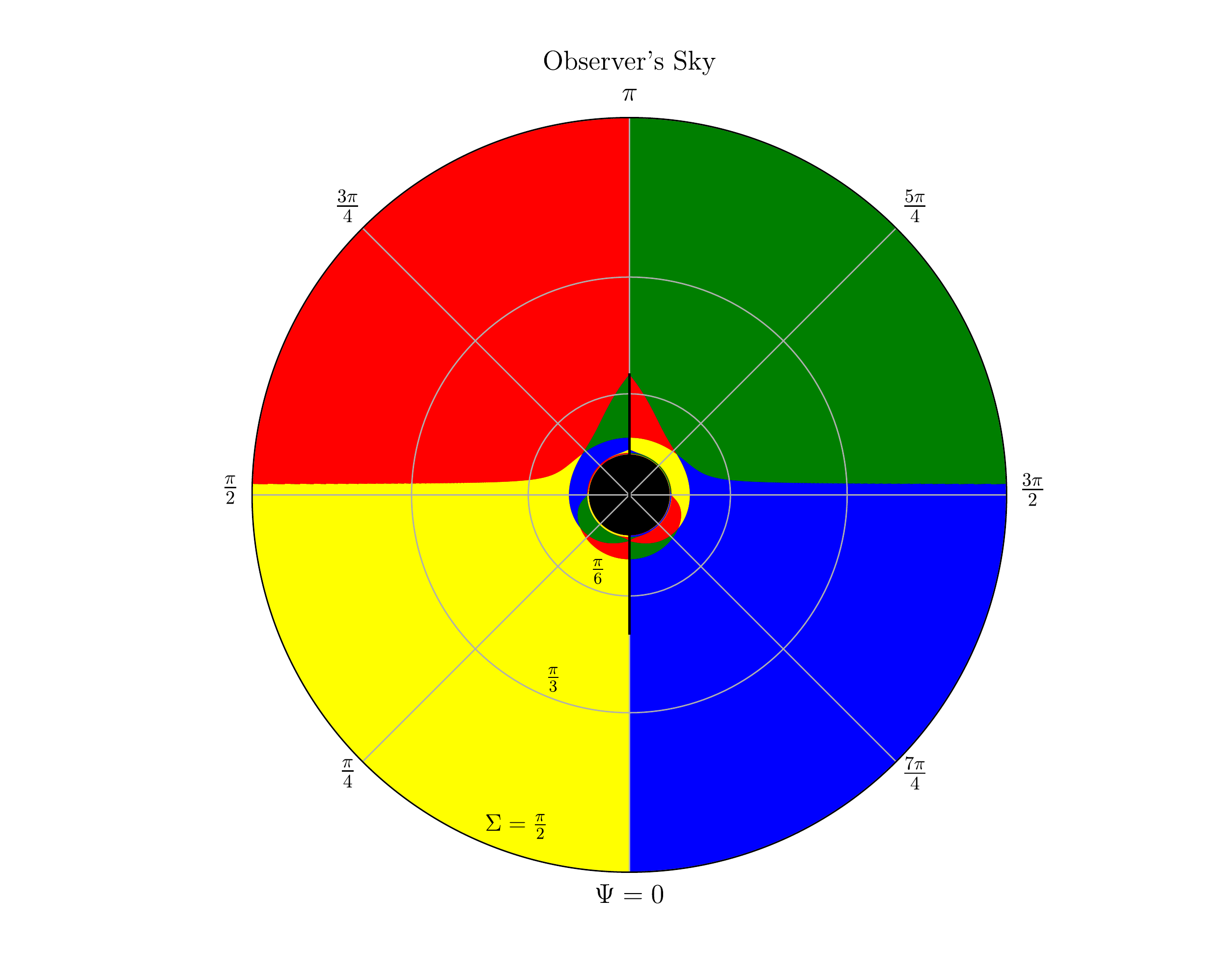} \\
\end{tabular}
\caption{Lens equation for the C-metric\cite{Frost2021} (top left), the charged
C-metric (top right), the C-de Sitter metric (bottom left) and the charged C-de
Sitter metric (bottom right) for $\Lambda=1/(200m^2)$, $e=m$ and $\alpha=1/(10m)$.
The observer is located at $r_{O}=8m$, $\vartheta_{O}=\pi/2$ and the sphere of light
sources is located at $r_{L}=9m$. The colour convention follows Refs.~\citenum{Frost2021} and \citenum{Bohn2015}
and is as follows: $0\leq\vartheta_{L}\leq\frac{\pi}{2}$: red/green; $\frac{\pi}{2}<\vartheta_{L}\leq\pi$:
yellow/blue $0\leq\varphi_{L}<\pi$: green/blue; $\pi\leq\varphi_{L}<2\pi$: red/yellow.
The black lines at $\Psi=0$ and $\Psi=\pi$ mark light rays that cross the axes at least once.}
\end{figure}
The lens map or lens equation maps images on the observer's sky back to the source
surface. This article is an extension of the work presented in Ref.~\citenum{Frost2021}
and therefore we will closely follow their definition of the lens map. To set up
the lens map we first distribute light sources on a sphere with radius $r_{O}<r_{L}$
in the region of outer communication. Then we shoot light rays backwards in time
from the position of the observer $(x_{O}^{\mu})$. Some of the light rays, but not
all, will intersect with the sphere of light sources. These geodesics form a map
from the celestial sphere of the observer to the sphere of light sources
\begin{eqnarray}
(\Sigma,\Psi)\rightarrow(\vartheta_{L}(\Sigma,\Psi),\varphi_{L}(\Sigma,\Psi)).
\end{eqnarray}
This is our lens equation. Now we have to calculate $\vartheta_{L}(\Sigma,\Psi)$
and $\varphi_{L}(\Sigma,\Psi)$ to obtain the lens map for the charged C-de Sitter
metric. For this purpose we set $(x_{i}^{\mu})=(x_{O}^{\mu})$ and insert (\ref{eq:CoM})
into the solutions of the equations of motion derived in Section~\ref{sec:EoM}. Due to the
symmetry of the charged C-de Sitter metric we can choose $t_{O}$ and $\varphi_{O}$
arbitrarily. Similarly $\lambda$ is only defined up to an affine transformation.
Therefore, to ease all following calculations we choose them such that $\lambda_{O}=0$,
$t_{O}=0$ and $\varphi_{O}=0$. \\
As first step towards calculating $\vartheta_{L}(\Sigma,\Psi)$ and $\varphi_{L}(\Sigma,\Psi)$
we calculate the Mino parameter $\lambda_{L}<\lambda_{O}$
\begin{eqnarray}
\lambda_L=\int_{r_{O}...}^{...r_{L}}\frac{\Omega(r_{O},\vartheta_{O})\mathrm{d}r'}{\sqrt{Q(r_{O})r'^4-r_{O}^2\sin^2\Sigma r'^2Q(r')}}.
\end{eqnarray}
Again for lightlike geodesics passing through a turning point we have to split the
integral and choose the sign of the root in the denominator in agreement with the
$r$ motion. Now we insert $\lambda_{L}$ in the appropriate solution for $\vartheta(\lambda)$
in Section~\ref{sec:theta} and obtain $\vartheta_{L}(\Sigma,\Psi)$. Next we count
the number of turning points $n$ of the $\vartheta$ motion. For this purpose we
calculate the Mino parameter up to the first turning point of the $\vartheta$ motion
$\lambda_{0}$. In the next step we calculate the difference of the Mino parameter
between two subsequent turning points of the $\vartheta$ motion $\Delta\lambda$.
Now we count how many turning points occur while $\lambda_{L}<\lambda_{n}=\lambda_{0}+n\Delta\lambda$.
Finally we calculate $\varphi_{L}(\Sigma,\Psi)$ from (\ref{eq:solphi}) as described
in Section~\ref{sec:phi}.\\
Fig.~4 shows plots of the lens equation in stereographic projection
for the C-metric (top left), the charged C-metric (top right), the C-de Sitter metric
(bottom left) and the charged C-de Sitter metric (bottom right) for $\Lambda=1/(200m^2)$,
$e=m$ and $\alpha=1/(10m)$. The observer is located at $r_{O}=8m$ and $\vartheta_{O}=\pi/2$.
The sphere of light sources is located at $r_{L}=9m$. The black lines at $\Psi=0$ and $\Psi=\pi$
mark lightlike geodesics that cross the string or the strut at least once. The black
circle in the centre of each image is the shadow of the black hole. The images clearly
show that the angular radius $\Sigma_{\mathrm{ph}}$ of the shadow decreases when $0<\Lambda$
and $0<e$. All four images basically show the same features up to a scaling. The
symmetry with respect to $\Psi=\pi/2$ and $\Psi=3\pi/2$ is clearly broken. Thus
in all four cases light rays ending parallel to the surface $\vartheta=\pi/2$ generally
come from light sources not located at $\vartheta=\pi/2$. However, the shape of
the features shown in the images is clearly symmetric with respect to the line marked
by $\Psi=0$ and $\Psi=\pi$. On the right side of the line of symmetry at the outer
boundary of each image we first have a region coloured in blue and green. In this
region we find images where the covered angle $0<\Delta\varphi<\pi$. These are images
of first order. Adjacent to these images, closer to the shadow we find a region
coloured in yellow and red. In this region we find images for which the light ray
covered the angle $\pi<\Delta\varphi<2\pi$. These are images of second order.
If we go closer to the shadow we also find images of third and fourth order. The
boundaries between the images of different orders mark the positions of the critical curves. On
the left side of the line of symmetry we find the same features, however, the ordering
of the colours is reversed. In addition we observe that at $\Psi=\pi$ images of second
order already occur further away from the shadow than for $\Psi=0$. This implies
that light rays passing close to the string cover the same angle $\Delta\varphi$
faster than light rays passing close to the strut.

\subsection{Redshift}
\begin{figure}\label{fig:Redshift}
\begin{tabular}{cc}
  C-Metric & Charged C-Metric \\[6pt]
  \includegraphics[width=65mm]{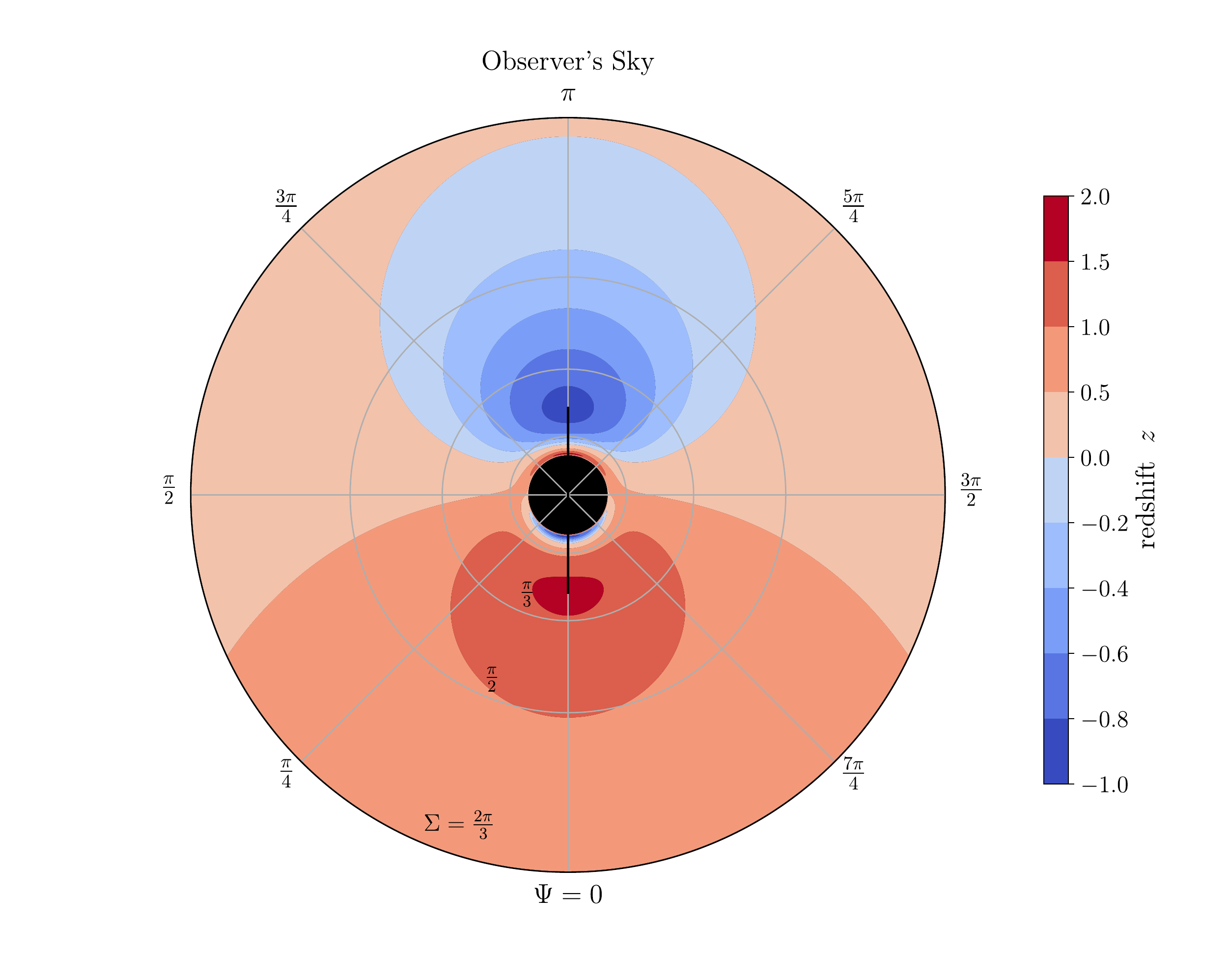} &   \includegraphics[width=65mm]{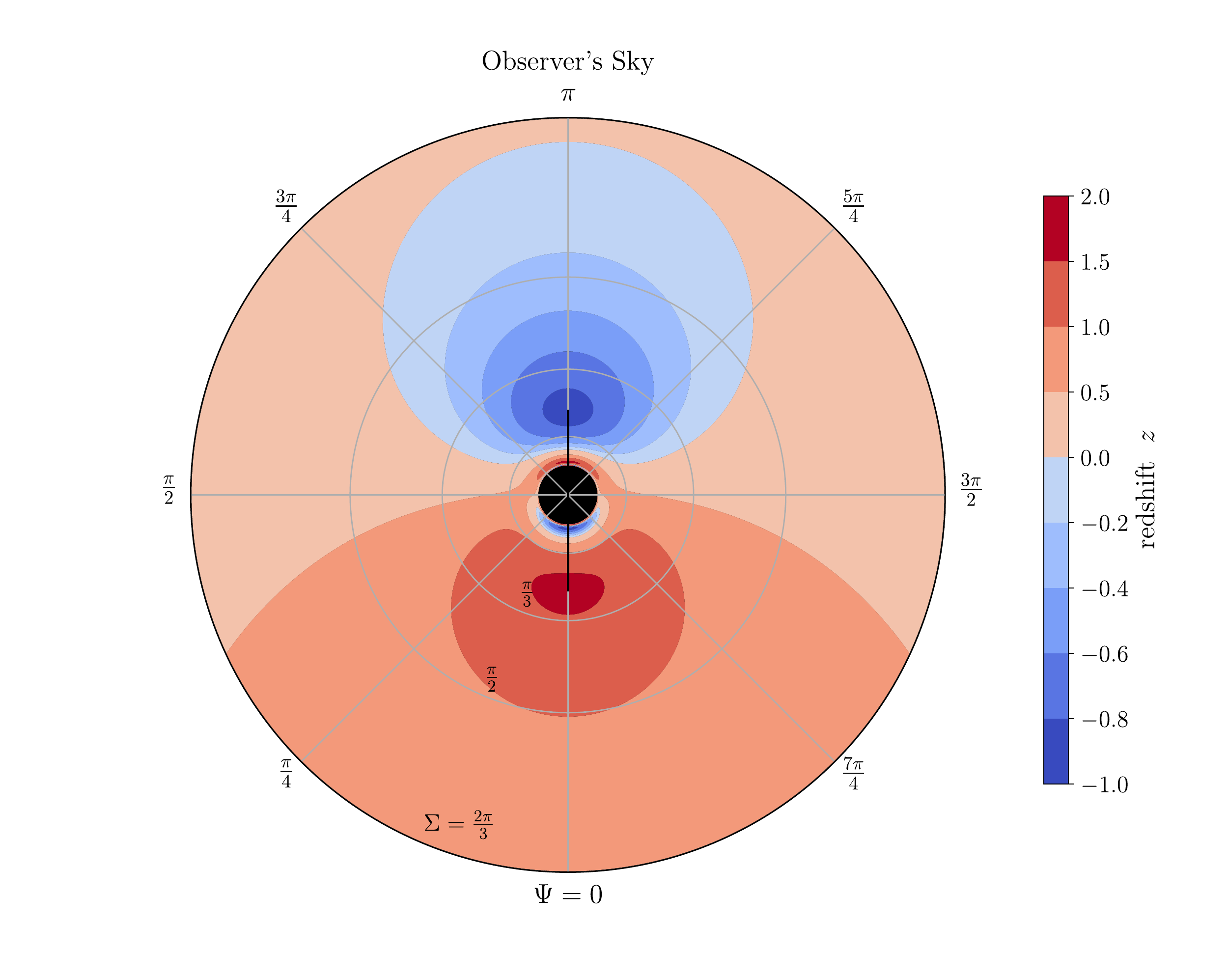} \\
  C-de Sitter Metric & Charged C-de Sitter Metric \\[6pt]
  \includegraphics[width=65mm]{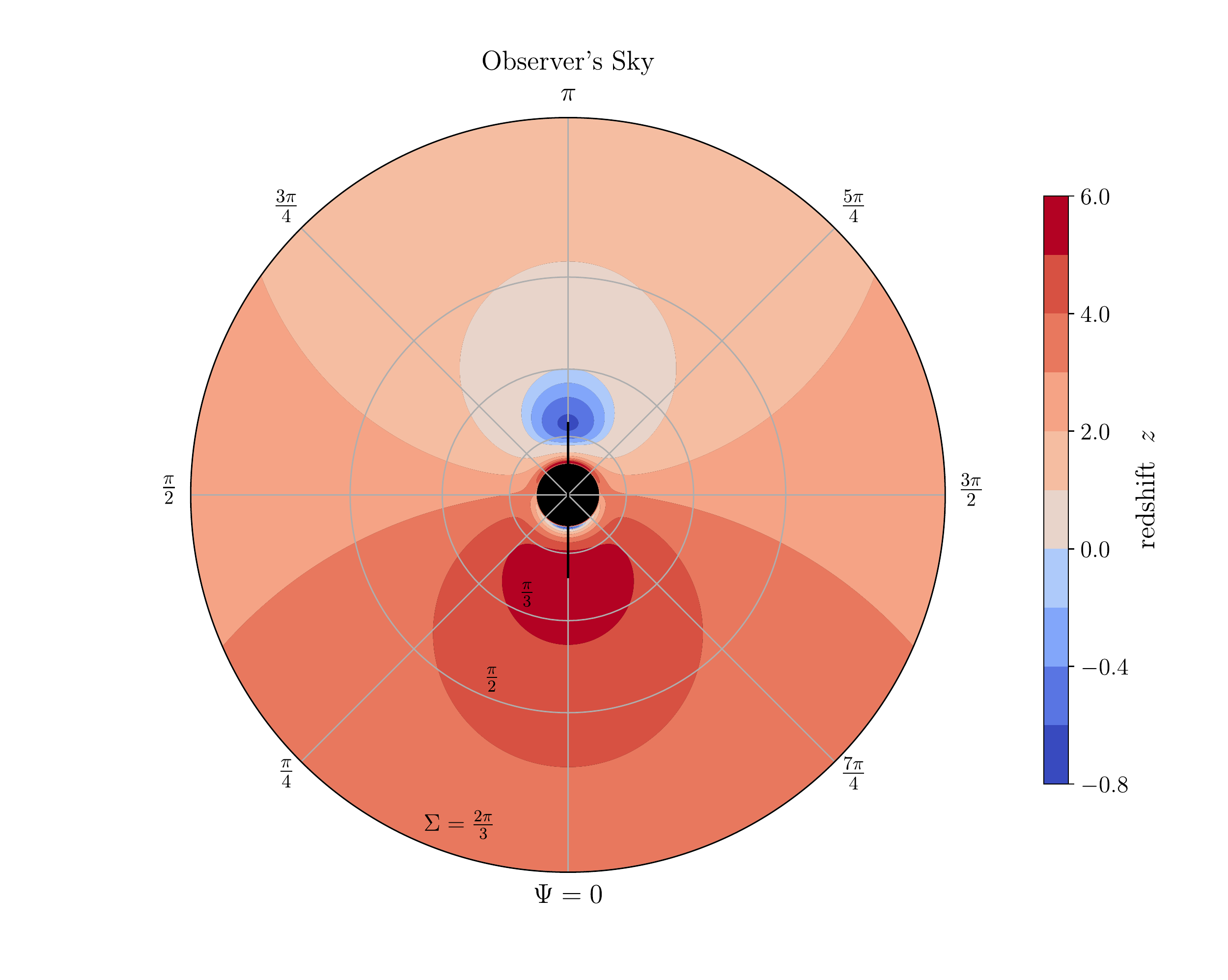} &   \includegraphics[width=65mm]{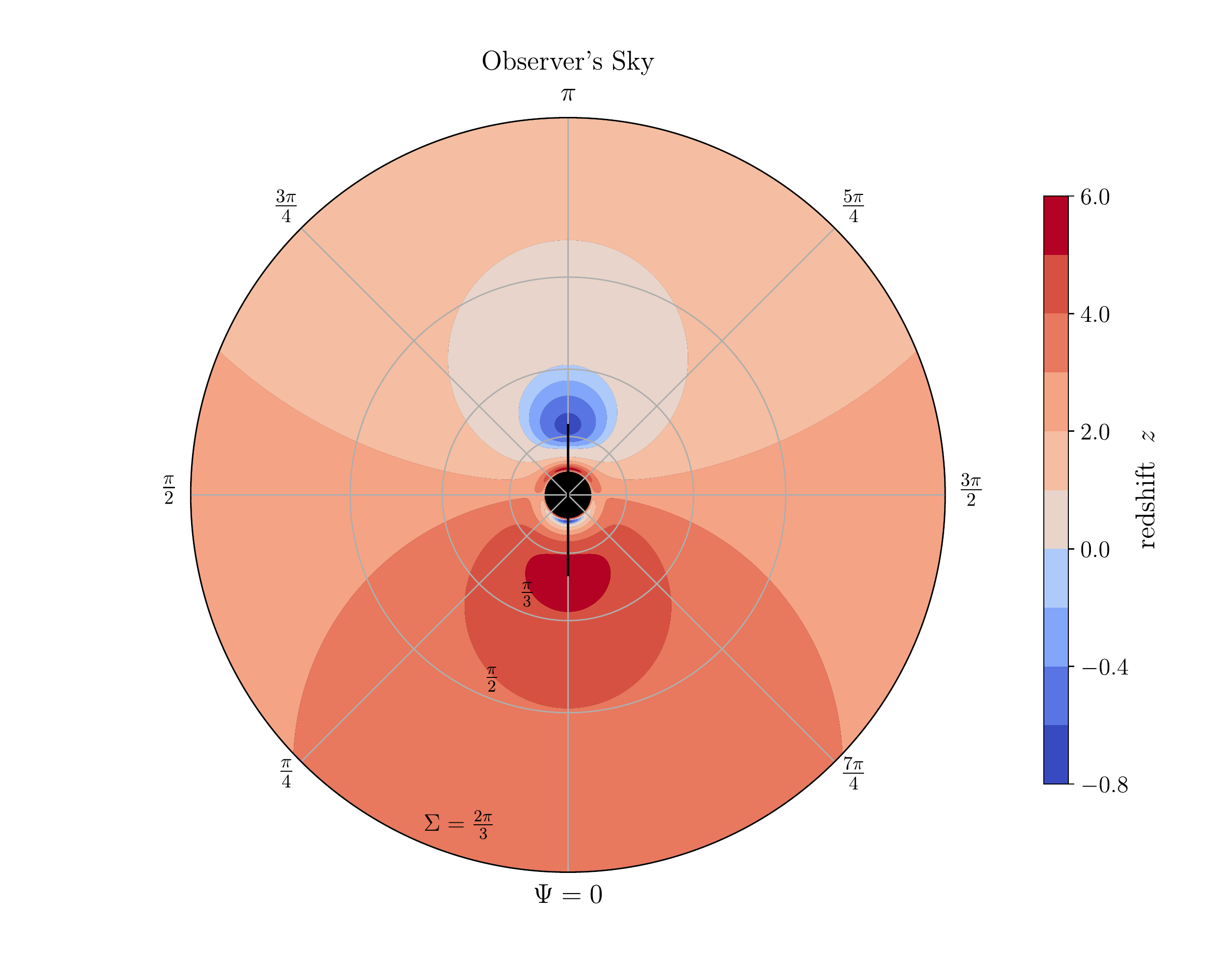} \\
\end{tabular}
\caption{Redshift maps for the C-metric\cite{Frost2021} (top left), the charged
C-metric (top right), the C-de Sitter metric (bottom left) and the charged C-de
Sitter metric (bottom right) for $\Lambda=1/(200m^2)$, $e=m$ and $\alpha=1/(10m)$.
The observer is located at $r_{O}=8m$, $\vartheta_{O}=\pi/2$ and the sphere of light
sources is located at $r_{L}=9m$. The black lines at $\Psi=0$ and $\Psi=\pi$ mark
light rays that cross the axes at least once.}
\end{figure}
The redshift $z$ of a light ray relates its energy at the time of emission by a light
source to its energy at the position at which it is detected by an observer. It
is directly accessible to observation via the frequency shift of atomic or molecular
absorption lines. We will now construct redshift maps. For this purpose we will
use the same settings and results we obtained from constructing the lens map in
Section \ref{sec:lensmap}. In this setting observer and light source are static
and the corresponding general redshift formula can be found in Ref.~\citenum{Straumann2013},
pp. 45. After inserting the metric coefficients $g_{tt}$ of the charged C-de Sitter
metric it reads\cite{Frost2021}
\begin{eqnarray}
z=\sqrt{\frac{\left.g_{tt}\right|_{x_{O}}}{\left.g_{tt}\right|_{x_{L}}}}-1=\sqrt{\frac{Q(r_{O})}{Q(r_{L})}}\frac{\Omega(r_{L},\vartheta_{L}(\Sigma,\Psi))}{\Omega(r_{O},\vartheta_{O})}-1.
\end{eqnarray}
Fig.~5 shows plots of the redshift maps for the same observer-source
geometry as for the lens maps in Fig.~4 In all four plots the outer region is dominated
by redshifts. The redshift has two maxima around $\Psi=\pi$ and at $\Psi=0$ (close
to the shadow). In addition in all images we observe a region of blueshifts centered
around $\Psi=0$. Another crescent shaped region of blueshifts can be found at $\Psi=\pi$
close to the shadow. What are now the effects of the cosmological constant $\Lambda$
and the electric charge $e$? Comparing the images with $\Lambda=0$ in the upper row
with the images with $\Lambda=1/(200m^2)$ in the lower row shows that when we turn
on the cosmological constant the redshift range shifts from $-1<z<2$ to $-0.8<z<6$,
respectively. The two blueshift areas move to lower latitudes closer to the centre of the
shadow. In addition the fraction of the images covered by blueshifts decreases.
The effects of the electric charge $e$ are less strongly pronounced. When we compare
the images on the left ($e=0$) to the images on the right ($e=m$) we observe that
in the top row the areas with blueshift move to slightly lower latitudes while they
still seem to cover roughly the same fraction of the image. However, comparing both
images in the lower row indicates that in the presence of a cosmological constant
turning on the electric charge leads to an increase of the fraction of the image
covered by blueshifts. In addition the areas of blueshifts shift to lower latitudes
but appear to be located at angular distances further away from the shadow.

\subsection{Travel Time}
\begin{figure}\label{fig:TT}
\begin{tabular}{cc}
  C-Metric/Charged C-Metric & C-de Sitter Metric/Charged C-de Sitter Metric \\[6pt]
  \hspace*{-0.7cm}\includegraphics[width=65mm]{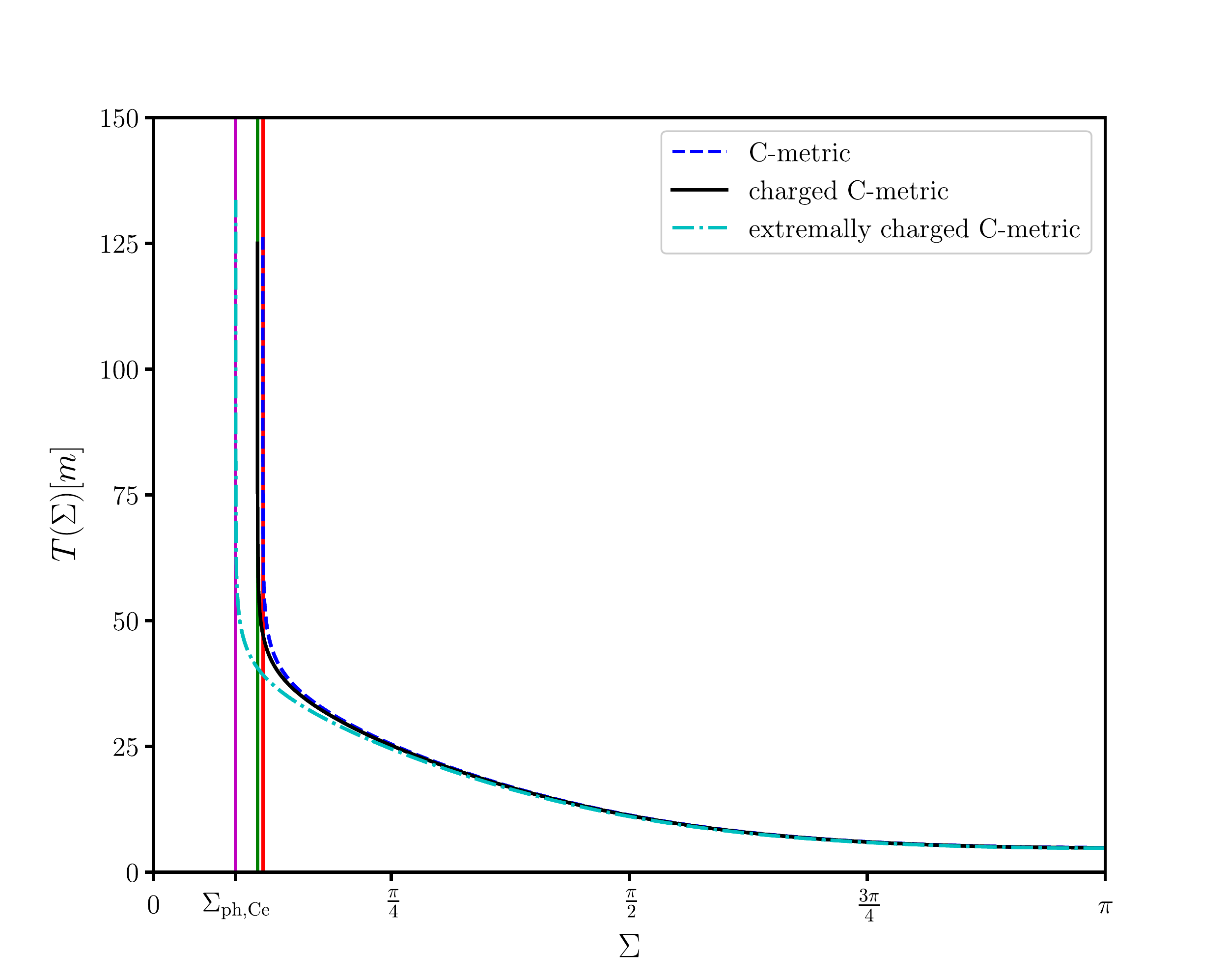} &   \includegraphics[width=65mm]{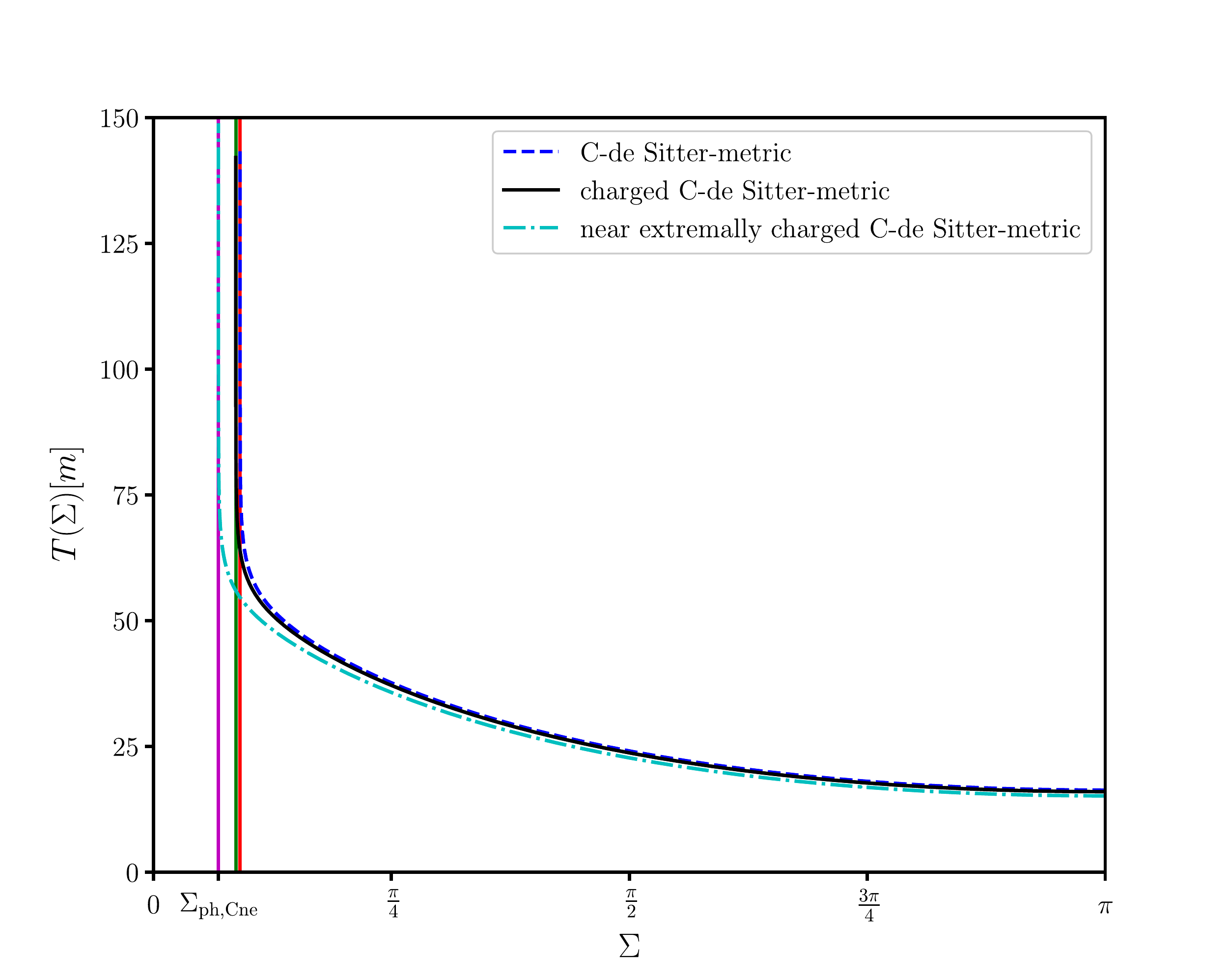} \\
\end{tabular}
\caption{Travel time $T(\Sigma)$ for the C-metric and the charged C-metric (left),
and the C-de Sitter metric and the charged C-de Sitter metric (right) for $\Lambda=1/(200m^2)$,
$e=m$ in the extremal and near extremal case, otherwise $e=m/2$ and $\alpha=1/(10m)$.
The observer is located at $r_{O}=8m$ and the sphere of light sources is located
at $r_{L}=9m$. $\Sigma_{\mathrm{ph},\mathrm{Ce}}$ and $\Sigma_{\mathrm{ph},\mathrm{Cne}}$ mark the position of the
angular radius of the shadow on the observer's celestial sphere for the extremally
charged C-metric and the near extremally charged C-de Sitter metric, respectively.}
\end{figure}
The travel time $T=t_{O}-t_{L}$ measures in terms of the time coordinate $t$ the
elapsed time between the emission of a light ray by a source at a time $t_{L}$ and
the detecion of the same light ray by an observer at the time $t_{O}$. We obtain
the travel time integral by inserting (\ref{eq:CoM}) and $t_{O}=0$ in (\ref{eq:Intt}).
We evaluate the travel time as described in Section~\ref{eq:solt}.
Fig.~6 shows the travel time for the (charged) C-metric (left) and the
(charged) C-de Sitter metric (right) for $\Lambda=1/(200m^2)$, $e=m/2$ in the regular
case, $e=m$ in the (near) extremal case and $\alpha=1/(10m)$. The observer is located
at $r_{O}=8m$ and the light source is located at $r_{L}=9m$. The left plot shows
that for $\Sigma>\pi/2$ in the C-metric and the charged C-metric the travel time
is roughly the same. When we turn on the cosmological constant this drastically
changes. In comparison to the (charged) C-metric the travel time gets significantly
longer. However, turning on the electric charge in presence of a positive cosmological
constant leads to a decrease of the travel time. In addition both plots indicate
that for $e=m/2$ the travel time close to $\Sigma_{\mathrm{ph}}$ is slightly shorter than
for $e=0$ while it is significantly longer for $e=m$.

\section{Summary and Implications for Observations}\label{sec:Conc}
In this article we extended the work presented for the C-metric in Ref.~\citenum{Frost2021}
to the charged C-de Sitter metrics. In the first part we discussed and solved the
equations of motion using elementary as well as elliptic functions and ellitpic
integrals. In the second part we used the derived analytical solutions to investigate
gravitational lensing in the charged C-de Sitter metrics. \\
How can we now use these results in combination with observational data to distinguish
between the different black hole spacetimes and in particular to measure the electric
charge of a black hole? \\
In the charged C-de Sitter metrics the shape of the shadow is always circular. Its
angular radius decreases with increasing $\Lambda$, $e$ and $\alpha$. However,
because the distance between the observer on Earth and the black hole lens is not
a priori known it alone cannot be used to determine the nature of the black hole.
The length of the travel time in particular close to the shadow is more characteristic
for each spacetime. Unfortunately in real astrophysical settings (multiple imaging
systems) we can only measure travel time differences leading to similar ambiguities
as for the shadow. However, the charged C-de Sitter spacetimes also admit two very
characteristic lensing features that distinguish them from their non-accelerating
counterparts ($\alpha=0$). The first characteristic is the breaking of symmetry
with respect to the equatorial plane on the celestial sphere of the observer in
the lens maps. Here, the most salient feature was the observation that images of
second order occur at larger (lower) angular distance from the shadow close to the
string (strut). The second characteristic is that the redshift $z$ is a function of
the coordinates on the observer's celestial sphere. The breaking of symmetry can
be tested by observing multiple images from the same light source gravitationally
lensed by a black hole. In such a system we can measure the position of the images
on the celestial sphere of the observer relative to the lens and then compare it
with theoretical predictions. Similarly although our construction of the redshift
map is highly idealised we may be able to find astrophysical systems similar to
this configuration. In such a system, we can measure the redshift of known emission
lines and construct a partial redshift map to determine if it is a function of
the coordinates on the observer's celestial sphere. Observing the symmetry breaking
and showing that the measured redshift is a function on the observer's celestial
sphere tells us with high certainty that the observed black hole is accelerating
and can be described by one of the charged C-de Sitter metrics. However, due to
the ambiguity introduced by the a-priori unknown distance between observer and black
hole lens we cannot use these measurements alone to determine $\Lambda$, $e$ and
$\alpha$.
Therefore, to correctly identify the nature of the spacetime describing the black
hole we observe, and having the chance to accurately measure the mass parameter
$m$, the cosmological constant $\Lambda$, the electric charge $e$ and the acceleration
parameter $\alpha$ we need to combine high accuracy measurements of the angular
diameter of the shadow, the redshift function on the observer's celestial sphere
and of the position of and the travel
time differences between two or more multiple images from the same source. Unfortunately
even with very high accuracy observations for most black holes the electric charge
$e$ is likely to be far to low to be measured and one may only be able to estimate
an upper limit.

\section*{Acknowledgments}
I would like to thank Volker Perlick for the helpful discussions. I acknowledge
financial support from QuantumFrontiers. I also acknowledge support from Deutsche
Forschungsgemeinschaft within the Research Training Group 1620 Models of Gravity.

\bibliographystyle{ws-procs961x669}
\bibliography{ChargedCdeSitterLensing}

\end{document}